\documentclass[11pt]{iopart}
\usepackage{comment}
\usepackage{enumerate}
\usepackage{amssymb}
\expandafter\let\csname equation*\endcsname\relax
\expandafter\let\csname endequation*\endcsname\relax
\usepackage{amsmath}
\usepackage{braket}
\usepackage{dsfont}
\usepackage{graphicx}
\usepackage[usenames,dvipsnames]{color}
\usepackage{tensor}
\usepackage{empheq}
\usepackage{capt-of}
\usepackage[normalem]{ulem} 
\usepackage{soul} %
\usepackage{MnSymbol}
\usepackage{cancel}

\usepackage[colorlinks,bookmarks=false,citecolor=NavyBlue,linkcolor=OliveGreen,urlcolor=blue]{hyperref}
\newcommand{\be}{\begin{equation}}
\newcommand{\ee}{\end{equation}}
\newcommand{\ba}{\begin{aligned}}
\newcommand{\ea}{\end{aligned}}
\newcommand{\bw}{\begin{widetext}}
\newcommand{\ew}{\end{widetext}}

\newcommand{\bea}{\begin{eqnarray}}
\newcommand{\eea}{\end{eqnarray}}

\def\doi{http://dx.doi.org/}

\begin{document}
\title{More on the operator space entanglement ($OSE$): R\'enyi $OSE$, revivals, and integrability breaking. }
\author{Vincenzo Alba}
\address{$^1$Dipartimento di Fisica dell'Universit\`a di Pisa and INFN, Sezione di Pisa, I-56127 Pisa, Italy}
\ead{vincenzo.alba@unipi.it}

\begin{abstract}

	We investigate the dynamics of the R\'enyi Operator Space Entanglement ($OSE$) entropies $S_n$ across 
	several one-dimensional integrable and chaotic models. 
	As a paradigmatic integrable system, we first consider the so-called rule $54$ chain. 
	Our numerical results reveal that the R\'enyi $OSE$ entropies of diagonal operators 
	with nonzero trace saturate at long times, in contrast with the behavior of von Neumann entropy. 
	Oppositely, the R\'enyi entropies of traceless operators exhibit logarithmic growth with time, 
	with the prefactor of this growth depending in a nontrivial manner on $n$. Notably, at long times, 
	the complete operator entanglement spectrum ($ES$) of an operator can be reconstructed from the spectrum 
	of its traceless part. We observe a similar pattern in the $XXZ$ chain, suggesting 
	universal behavior. Additionally, we consider dynamics in nonintegrable deformations of the  $XXZ$ chain. 
	Finite-time corrections do not allow to access the long-time behavior of the  
	von Neumann entropy. On the other hand, for $n>1$ the growth of the entropies is  
	milder, and it is compatible with a sublinear growth, at least for operators associated with global 
	conserved quantities. 
	Finally, we show that in finite-size integrable systems, $S_n$ exhibit strong revivals, 
	which are washed out when integrability is broken. 

\end{abstract}

\maketitle

\section{Introduction}
\label{sec:intro}

\begin{figure}[t]
\begin{center}
\includegraphics[width=.7\textwidth]{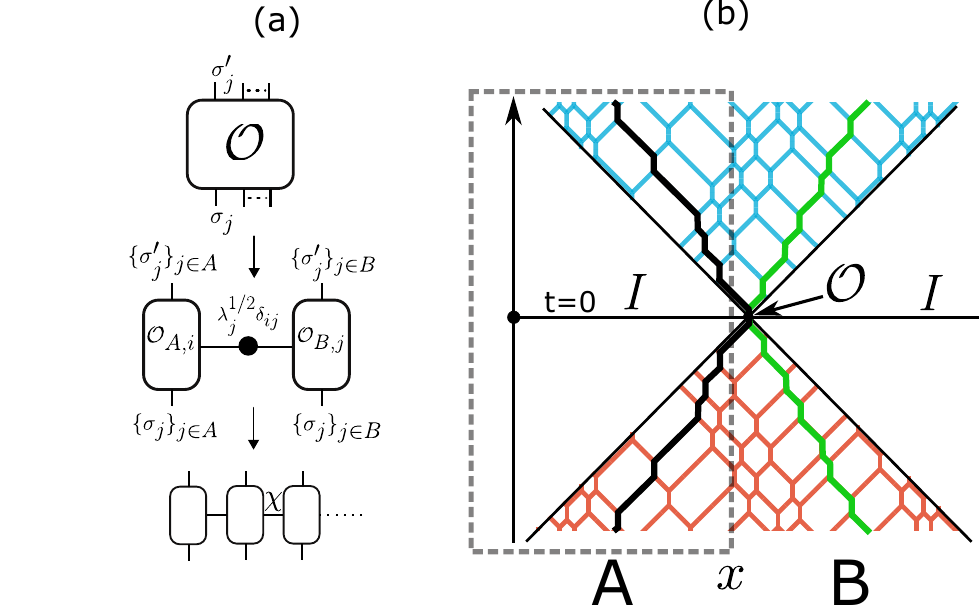}
\caption{Setup considered in this work. Panel (a) shows the Matrix Product Operator 
	($MPO$) representation of a generic operator $\mathcal{O}$. $\mathcal{O}$ is represented 
	as a box, the legs being the row and column indices $\sigma_j$ and $\sigma_{j}'$. 
	Here we restrict ourselves to spin-$1/2$ systems with $\sigma_j,\sigma_j'=\uparrow,\downarrow$. $\mathcal{O}$ is represented as a $2^L\times 2^L$ matrix, with $L$ the total 
	number of sites. By dividing the system in two complementary regions $A$ and $B$, 
	one can always perform a Schmidt decomposition of the 
	operator (cf.~\eqref{eq:schmidt}) in terms of two orthonormal bases $\mathcal{O}_{A,i}$ and $\mathcal{O}_{B,j}$ for $A,B$ and the singular values $\sqrt{\lambda_j}$. By iterating 
	the Schmidt decomposition within $A$ and $B$ and keeping at each step the largest 
	$\chi$ (bond dimension) singular values only, one 
	obtains a compressed $MPO$ representation for $\mathcal{O}$. (b)
We study the Heisenberg evolution 
	$\hat{\mathcal{O}}(t)=e^{iHt/\hbar}\hat{ \mathcal{O}}e^{-iHt/\hbar}$ of a 
	local operator $\hat{\mathcal{O}}$ acting nontrivially at the center of a one dimensional system. 
	Away from the center the operator acts as the identity. 
	As time progresses, the support of the operator spreads, forming a double lightcone. For diagonal 
	operators, the upper and lower parts of the lightcone coincide.  
	We study the dynamics of the R\'enyi Operator Space Entanglement ($OSE$) entropies $S_n(\hat{\mathcal{O}})$ 
	($n\in\mathbb{N}$) of region $A$, starting at position $x$. Here we fix $x=0$, meaning that 
	$A$ is the half chain starting from the operator insertion. In the rule $54$ chain 
	the operator spreading is understood in terms of interacting left and right movers (solitons). 
}
\end{center}
\label{fig0:cartoon}
\end{figure}

The entanglement growth in typical out-of-equilibrium dynamics~\cite{calabrese-2005,fagotti2008evolution,
kim2013ballistic,liu2014entanglement,alba2017entanglement,bertini2022growth}  
significantly hinders the simulation 
of out-of-equilibrium quantum many-body systems with state-of-the-art 
numerical methods, such as the time-dependent Density Matrix Renormalization 
Group~\cite{schollwoeck2011the,paeckel2019time} ($tDMRG$). 
Early research into $tDMRG$ suggested that instead of simulating 
the dynamics of the full quantum state, it could be convenient to focus 
on the Heisenberg dynamics of operators. The Heisenberg evolution of  
an operator $\hat{\mathcal{O}}$ is expressed as 
\begin{equation}
	\hat{\mathcal O}(t)={\hat U}^\dagger(t)\hat{\mathcal O}(0)\hat U(t),\quad \hat{U}(t):=e^{-i Ht/\hbar}, 
\end{equation}
with $H$ the Hamiltonian of the system. 
Here we restrict ourselves to one-dimensional lattice models (spin chains), considering a generic 
local operator 
$\hat{\mathcal O}$ that acts nontrivially only at the center of the chain, and is the identity 
elsewhere (see Fig.~\ref{fig0:cartoon}). 
It is important to quantify the complexity of the operator growth,  
particularly by constructing the time-dependent Matrix Product Operator~\cite{paeckel2019time} 
$(MPO)$ of $\hat{\mathcal O}$. 

The construction of $MPO$s is illustrated in Fig.~\ref{fig0:cartoon} (a). Let us consider a 
generic operator $\mathcal{O}$ living on a system of $L$ sites. We focus on 
spin-$1/2$ systems. The operator $\mathcal{O}$ is represented as 
a $2^L\times 2^L$ matrix (see the box in Fig.~\ref{fig0:cartoon} (a)) 
in the standard computational basis $|\sigma_1,\sigma_2,\dots,\sigma_L\rangle$, with 
$\sigma_j=\uparrow,\downarrow$. 
Let us bipartite the system as $A\cup B$ (see Fig.~\ref{fig0:cartoon}). 
It is always possible to perform a Schmidt decomposition of $\hat{\mathcal O}$ as 
\begin{equation}
	\label{eq:schmidt}
	\frac{\hat{\mathcal O}(t)}{\sqrt{\mathrm{Tr}
	({\hat{\mathcal O}}^\dagger\hat{\mathcal O})}}=
	\sum_i\sqrt{\lambda_i}\hat {\mathcal O}_{A,i}\otimes\hat {\mathcal O}_{B,i},
\end{equation}
where  $\hat{\mathcal O}_{A/B,i}$ are two orthonormal sets of operators with 
support in $A$ and $B$, satisfying $\mathrm{Tr}({\hat{\mathcal O}_{A/B,i}}^\dagger
\hat{\mathcal O}_{A/B,j})=\delta_{ij}$. In~\eqref{eq:schmidt} $\sqrt{\lambda_i}$ are 
the so-called Schmidt values. The decompositon in~\eqref{eq:schmidt} (see Fig.~\ref{fig0:cartoon}) is 
exact. By keeping only the largest $\chi$ Schmidt numbers, 
one obtains a truncated representation of $\hat{\mathcal{O}}$. 
Moreover, by iterating the Schmidt decomposition within the two regions $A,B$, 
keeping at each step the largest $\chi$ Schmidt values, one arrives at a compressed $MPO$ 
representation of $\hat{\mathcal{O}}$. In contrast with the original representation 
of the operator, which depends on $2^{2L}$ complex numbers, the $MPO$ is determined 
by $d^2\chi L$ numbers, where $d=2$ is the dimension of the local Hilbert space. 
Here we consider operators that at $t=0$ act nontrivially only at the center of the 
chain, and are the identity elsewhere. The $MPO$ of such operators at $t=0$ is exact with bond 
dimension $\chi=1$. Understanding the growth of the bond dimension $\chi$  
as a function of time is crucial to quantify the effectiveness  of 
the $MPO$ decomposition. 

Quite generically, at $t>0$ a double lightcone spreads from the 
center of the chain (see Fig.~\ref{fig0:cartoon} (b)),  as the operator  support, i.e., the sites of the chain 
where the operator acts nontrivially, expands with time. This is typically accompanied by the growth of 
$\chi$ of the operators that live within the lightcone. Based on exact results in noninteracting systems, 
Ref.~\cite{prosen2008is} conjectured that there is a significant  
difference between integrable and nonintegrable dynamics: 
In integrable systems the bond dimension necessary for a  fixed fidelity  approximation of 
a local operator grows only polynomially with time, whereas nonintegrable Hamiltonians  
lead to an exponential growth of $\chi$. 

The operator space entanglement ($OSE$) entropies~\cite{zanardi2001entanglement} 
$S_n(\hat{\mathcal O})$ provide the means to quantify the growth of $\chi$ with time, and to 
address these issues. 
%
%
%
In the following we mostly consider the situation with $A$ and $B$ being the two 
halves of the chain. We also fix $x=0$ (see Fig.~\ref{fig0:cartoon}), placing the boundary 
between $A$ and $B$ at the operator insertion. The R\'enyi $OSE$ entropies $S_n$ 
and the von Neumann one $S_1$ are defined as 
\begin{equation}
	\label{eq:renyi-vn} 
	S_n(\hat {\mathcal O})=\frac{1}{1-n}\ln\Big(\sum_i\lambda^n_i\Big),
	\quad S_1=-\sum_i\lambda_i\ln(\lambda_i). 
\end{equation}
Now, according to the conjecture in Ref.~\cite{prosen2008is}, integrable dynamics lead to 
at most a logarithmic growth of $S_n$ with time, whereas nonintegrable ones are expected to give linear growth of 
$S_n$. This scenario was first verified in noninteracting models~\cite{prosen2007operator,pizorn2009operator,dubail2017entanglement}. 
Recently, it has been shown~\cite{alba2019operator,alba2021diffusion} that the von Neumann 
$OSE$ entropy can grow at most logarithmically with time in the rule $54$ chain~\cite{bobenko1993on,buca2021rule}, 
which has been studied intensively~\cite{gopalakrishnan2018operator,gopalakrishnan2018hydrodynamics,
prosen2016integrability,gopalakrishnan2018,buca2019exact,rowlands2018noisy} 
as a toy model of interacting integrable systems. 
Precisely, based on previous results in Ref.~\cite{klobas2019time}, 
 Ref.~\cite{alba2019operator} provided an algorithm to construct 
the $MPO$ for a generic local operator $\hat{\mathcal O}(t)$. Crucially, 
the bond dimension $\chi$ grows as $t^2$, which implies that $S_n$  
can grow only logarithmically with time. Moreover, Ref.~\cite{alba2019operator} 
numerically showed that a similar logarithmic bound holds in Bethe 
ansatz solvable systems, such as the spin-$1/2$ Heisenberg $XXZ$ chain.  
The linear-growth scenario for nonintegrable dynamics  is supported by 
simulations in random unitary circuits, for which a so-called ``membrane'' picture for entanglement 
spreading applies~\cite{nahum2017quantum,jonay2018coarse,chan2018solution,vonkeyserlingk2018operator,zhou2020entanglement}. 
Exact results in solvable chaotic quantum 
circuits~\cite{bertini2020scrambling,bertini2020operator,bertini2020operatorentanglement,claeys2020maximum,
foligno2023temporal,rampp2023from,rampp2024entanglement} are also in agreement with the membrane picture.  

Several open questions remain to be investigated. For instance, the  mechanism 
underlying the growth of the $OSE$ has not been clarified yet. While Ref.~\cite{alba2019operator} 
showed logarithmic growth for the $OSE$ entropies in integrable systems, the prefactor of the 
logarithmic growth is not understood, except for some 
simple operators~\cite{alba2021diffusion}. More importantly, the relationship 
between transport properties and operator spreading, if any, is yet to be unveiled. 
Still, it has been argued in Ref.~\cite{alba2021diffusion}  
that the growth of the $OSE$ of some simple operators reflects the diffusive fluctuations 
of the operator front~\cite{lopez2021operator}. Furthermore, while a lot of attention has been 
devoted to the von Neumann $OSE$ entropy, much less is know about the R\'enyi entropies, or the 
operator entanglement spectrum ($ES$) formed by the squared Schmidt values 
$\{\lambda_i\}$ (cf.~\eqref{eq:schmidt}). 
Clearly, it is of utmost importance to understand the scaling of $OSE$ in nonintegrable Hamiltonians. 
Unfortunately, this is a daunting task from the numerical perspective, as $tDMRG$ 
methods rely on the mild grow of the bond dimension with time. Interestingly, it has been suggested~\cite{muth2011dynamical} 
(see also Section~\ref{sec:bound}) that even in  nonintegrable systems the $OSE$ 
entropies of operators linked with global conservation laws could 
exhibit logarithmic growth with time, in contrast with the membrane-picture scenario.

Here we address some of these open questions focusing on diagonal operators. We first consider the dynamics  of 
the R\'enyi  entropies in the rule $54$ chain. We show that there is a crucial distinction to be made 
between traceless operators and operators with nonzero trace. Precisely, $S_n$ grow logarithmically 
for any $n$ for traceless diagonal operators. The prefactor of the logarithmic growth 
depends on $n$ in a nontrivial manner, and it is different from that governing the logarithmic growth of  
operators in noninteracting systems~\cite{dubail2017entanglement}, which can be obtained with 
Conformal Field Theory ($CFT$) techniques. 
We also investigate  the distribution of the operator entanglement spectrum levels, which shows 
marked differences with that obtained in Conformal Field Theories~\cite{calabrese2008entanglement}. 
Oppositely, the entropies $S_n$ of operators with nonzero trace 
grow logarithmically with time only for $n=1$, in agreement with previous results~\cite{alba2019operator,alba2021diffusion}. 
For $n\ne1$,  $S_n$ saturate at long times, the saturation values being fully determined by the trace of the operator. 
Interestingly, we show that the complete entanglement spectrum of an operator can be determined from the spectrum of 
its traceless part. We should mention that while the same results were presented as an approximation  
in Ref.~\cite{jonay2018coarse}, here we provide evidence that they become asymptotically exact in the long time 
limit. 
To address the universality of our results, we consider the spin-$1/2$ $XXZ$ chain, which is a paradigmatic 
interacting integrable system. 
The same scenario uncovered for the rule $54$ chain applies: 
Both the R\'enyi and the von Neumann entropies of traceless operators grow logarithmically with time for any $n$. 
Interestingly, within the accuracy of our $tDMRG$ simulations, the dependence on $n$ of the prefactors of the logarithmic 
growth is the same as in the rule $54$ chain, suggesting the same 
underlying mechanism for operator spreading in the two models. 
The R\'enyi $OSE$ entropies of operators with nonzero trace saturate at long time. 

Next we consider the effects of breaking integrability, focusing on the $XXZ$ chain with 
next-nearest-neighbor interactions. 
First, we observe that $S_n$ exhibit strong finite-time corrections. This 
makes challenging to determine the asymptotic behavior  of the von Neumann entropy at long times. 
On the other hand, our numerical results for $S_n$ with $n>1$ are consistent with a 
quite mild growth in agreement with the bound of Ref.~\cite{muth2011dynamical}. 
Finally, we investigate revival effects in finite-size chains. 
We show that for dynamics in the $XXZ$ chain, which is integrable, the  entropies  
exhibit revivals at times $t\approx L/2$. Revivals correspond to localized 
quasiparticles originated at the center of the chain 
that re-enter subsystem $A$  after bouncing back at the boundary of the system. Interestingly, 
in the presence of integrability-breaking interactions, the revivals are washed out, which 
reflects the absence of well-defined quasiparticles. 

The manuscript is organized as follows. In Section~\ref{sec:intro} we introduce the rule $54$ chain 
(in Section~\ref{sec:rule-54}) and the deformed $XXZ$ chain (in Section~\ref{sec:rule-54}). In particular, 
in Section~\ref{sec:rule-54}  we discuss some general results for operator spreading in integrable systems. 
In Section~\ref{sec:bound} we review exact results for the $OSE$ entropies. Specifically, in Section~\ref{sec:muth} 
we focus on Ref.~\cite{muth2011dynamical}, which provided an exact bound between R\'enyi  entropies of an operator $\hat{\mathcal{O}}$ and 
its  Infinite Temperature Auto Correlator ($ITAC$) $\langle \hat{\mathcal{O}}(t)\hat{\mathcal{O}}(0)\rangle_{T=\infty}$, with $\langle\cdot\rangle_{T=\infty}$ denoting the average over the infinite-temperature ensemble. In Section~\ref{sec:traceful}, following Ref.~\cite{jonay2018coarse}, we discuss 
the relationship between the operator entanglement spectrum of traceless operators  and that of operators with 
nonzero trace. In Section~\ref{sec:diffusion} we discuss the results of Ref.~\cite{alba2021diffusion}. 
In Section~\ref{sec:renyi-54} we present numerical $tDMRG$ data for the dynamics of diagonal operators in 
the rule $54$ chain. Precisely, in Section~\ref{sec:trace} we show that for operators with 
nonzero trace only the von Neumann entropy grows logarithmically, whereas the R\'enyi entropies saturate at long times. 
In Section~\ref{sec:noncft} we show that the R\'enyi entropies of traceless operators grow logarithmically. 
The dynamics of the operator entanglement spectrum is discussed in Section~\ref{sec:spetrum-54}. 
In Section~\ref{sec:xxz-num} we focus on the $XXZ$ chain. In Section~\ref{sec:trace-xxz} we compare the 
behavior of the R\'enyi entropies of traceless and operators with nonzero trace, showing that 
the same scenario discussed for the rule $54$ chain in Section~\ref{sec:renyi-54} applies. Section~\ref{sec:int-break} 
is devoted to the effects of integrability breaking. In Section~\ref{sec:revivals} we 
discuss revivals in finite-size integrable and nonintegrable systems. We conclude in Section~\ref{sec:concl}.

\section{Models \& setup: Operator spreading in the rule $54$ chain and the $XXZ$ chain}
\label{sec:intro}

Here we introduce the two models that we will consider in this work, namely the so-called rule $54$ chain 
and a nonintegrable deformation of the $XXZ$ chain. Specifically, in Section~\ref{sec:rule-54} we discuss 
extensively the operator spreading in the rule $54$ chain, as it is paradigmatic of generic 
integrable systems. In Section~\ref{sec:xxz} we introduce the spin-$1/2$ Heisenberg $XXZ$ chain with 
next-to-nearest-neighbor interactions, which allows to investigate both integrable and nonintegrable 
dynamics.

\subsection{Rule $54$ chain}
\label{sec:rule-54}

The Hilbert space of the rule $54$ chain~\cite{bobenko1993on} is 
that of a system of classical bits $s_x=0,1$. The system lives on a 
chain of $L$ sites. The dynamics is generated by a three-site unitary gate $U_{x}$ 
acting as
\begin{equation}
	\label{eq:gate}
	U_x=|s_{x-1},s'_x,s_{x+1}\rangle\langle s_{x-1}s_x,s_{x+1}|,
\end{equation}
where $s'_x=s_{x}+s_{x-1}+s_{x+1}-s_{x-1}s_{x+1}\,\mathrm{mod}\,\,2$. Precisely, one has 
\begin{multline}
	\label{eq:gate-1}
	U_x=|101\rangle\langle 111|+
|100\rangle\langle 110|+
|111\rangle\langle 101|+
|110\rangle\langle 100|\\+
|001\rangle\langle 011|+
|010\rangle\langle 010|+
|011\rangle\langle 001|+
|000\rangle\langle 000|. 
\end{multline}
As it is clear from~\eqref{eq:gate} and~\eqref{eq:gate-1}, 
$U_x$ flips the bit at position $x$ if at least one of its neighboring bits at $x-1$ and $x+1$ 
is in the state $1$. 
Any bit configuration is evolved by applying $U_x$  in a ``brick-wall'' 
fashion as  $U=\prod_{\mathrm{even}\, x}\prod_{\mathrm{odd}\,x}U_x$. Time is 
discrete in units of $1$. Clearly, Eq.~\eqref{eq:gate} maps classical bit strings 
into classical bit strings. 
The rule $54$ chain possesses well-defined quasiparticles (see Fig.~\ref{fig0:cartoon}), which is the key property of
generic integrable systems. Quasiparticles are emergent left/right moving solitons. Precisely, 
solitons are formed by pairs of adjacent bits that are in the $1$ state. Depending on 
the parity of the site of the leftmost $1$ of the pair, they are left or right moving solitons~\cite{klobas2019time}. 
The velocity of the solitons is $v=\pm1$. Solitons are interacting, undergoing pairwise elastic scattering. 
The scattering is implemented as a Wigner time delay~\cite{wigner1955lower}. Precisely, 
when a left and a right moving soliton meet at the same site, they merge giving rise to 
the configuration $\cdots 010\cdots$. The solitons remain ``bound'' together for a unit 
of time, before re-emerging as distinct left and right movers (several scattering processes are 
shown in Fig.~\ref{fig0:cartoon}). 
Again, this behavior is reminiscent of generic integrable models, which exhibit factorized 
two-body elastic scattering. The major difference is that in generic integrable models 
the quasiparticles possess a nontrivial dispersion, implying that their energies and 
velocities  depend on a real parameter $\lambda$, which labels the different quasiparticles. 
It is important to stress that there is a mapping between the soliton basis and the 
standard bit basis. The mapping can be represented as an $MPO$. By applying the $MPO$  
on a generic string of bits,  one obtains the corresponding configuration of left and right movers. 
Crucially, the mapping is local, because 
to determine whether at a generic site $x$ there is a left of right moving soliton, only the 
bit configuration of the neighboring sites is needed. This implies that the mapping can be represented as 
an $MPO$ with finite bond dimension $\chi$ (see Ref.~\cite{alba2019operator} and~\cite{alba2021diffusion} for 
the explicit mapping). The bond dimension of the $MPO$ that encodes the mapping  does not depend on time, and 
cannot affect the leading behavior of the 
operator entanglement in the large time limit. 

Let us now discuss operator spreading in the rule $54$ chain. First, let us observe that the identity operator 
is 
\begin{equation}
	\label{eq:id}
	\mathds{1}=\prod_x (|0\rangle\langle0|+|1\rangle\langle1|)_x, 
\end{equation}
where the subscript $x$ means that the operator lives in the Hilbert space at site $x$. Upon expanding 
the product in~\eqref{eq:id}, one obtains the equal-weight superposition of all the terms with an arbitrary number 
of projectors $|0\rangle\langle0|$ and $|1\rangle\langle1|$. The identity commutes with the evolution operator, 
meaning that the equal-weight superposition is mapped onto itself under the dynamics. 
The $MPO$ representing the identity has bond dimension $\chi=1$. This is evident from 
the fact that in the bit basis, irrespective of the local projector present at site $x$, the neighboring 
sites at $x+1$ and $x-1$ can have both $|0\rangle\langle0|$ or $|1\rangle\langle1|$. This implies that 
$S_n(\mathds{1})=0$ for any $n$, and at any time. 

The behavior changes dramatically if an operator ${\mathcal O}$ 
is inserted at the center of the system, because the bond dimension  of the $MPO$ 
of the operator starts growing with time. Specifically, a 
double lightcone forms (see 
Fig.~\ref{fig0:cartoon} (b)), as information propagates away from the operator insertion at finite velocity. The $MPO$ 
of the operator has bond dimension $\chi=1$ outside of the lightcone, where the operator remains the identity, 
whereas inside the lightcone its bond dimension grows. The reason is that the operator insertion 
puts nonlocal constraints on the motion of the left and right movers. 
To illustrate that,  we now focus on the spreading of diagonal operators, for which the double-lightcone 
is symmetric, i.e., its upper part coincides with its lower one. Specifically 
we consider the local projector operator $P_z$ acting at the center of the system, and 
$S^z_{L/2}$, defined as 
\begin{equation}
	\label{eq:Pz-Sz}
	P_z:=\frac{1}{2}\left(\mathds{1}+\sigma_{L/2}^z\right)=|1\rangle\langle1|, 
	\quad S_{L/2}^z:=\sigma_{L/2}^z,  
\end{equation}
with $\sigma_x^{z}$ the Pauli matrix acting at site $L/2$. 
Notice that while $S^z$ is a traceless operator, $P_z$ has a 
nonzero trace. 
Operator spreading is better understood in the soliton basis~\cite{alba2019operator}. 
To represent $P_z$ in the basis of left and rigth movers, it is convenient to expand 
$\mathds{1}_{L/2-1}P_z\mathds{1}_{L/2+1}$ as 
\begin{equation}
	\label{eq:Pz}
\mathds{1}_{L/2-1}P_z\mathds{1}_{L/2+1}=|010\rangle\langle010|+|011\rangle\langle011|+
|110\rangle\langle110|+|111\rangle\langle111|. 
\end{equation}
The first term in~\eqref{eq:Pz} creates a pair of scattering solitons (as depicted in Fig.~\ref{fig0:cartoon} (b)), 
whereas the second and third one 
creates a left and a right moving soliton at the center, respectively. Upon applying the evolution operator~\eqref{eq:gate} 
the solitons created by the insertion of the operator propagate, scattering with the background 
solitons away from the center. This means that at a generic 
time $t$ any allowed soliton configuration in the light cone has to contain the solitons that were 
created by the operator insertion at $t=0$, although their positions are shifted by the scatterings with 
the background solitons (see Fig.~\ref{fig0:cartoon} (b)). All the \emph{allowed} soliton configurations at time $t$ 
have the same amplitude. One can show that the bond dimension of the evolved $MPO$ is determined by the number of steps 
to decide whether a generic soliton configuration in the ligthcone is allowed 
or not~\cite{klobas2019time,alba2019operator}. Equivalently, this is the number of steps needed to identify 
the positions at time $t$ of the solitons that emerged from the center of the chain, and it grows as $t^2$. 
Importantly, the fact that  $\chi={\mathcal O}(t^2)$ provides only a bound on the growth of the $OSE$ entropies. Indeed, 
by assuming that all the $\chi$ Schmidt values in~\eqref{eq:schmidt} are the same, one obtains 
$S_1\propto 2\ln(t)$. 
On the other hand, Ref.~\cite{alba2019operator} provided numerical evidence that $S_1\propto \ln(t)$ 
for the operator $S^z$ and $S_1\propto 1/2\ln(t)$ for $P_z$~\cite{alba2021diffusion}. This means that the exact $MPO$s  
describing the dynamics of local operators contain too much information 
and  can be further compressed. The reason why this is possible has not been clarified yet. 

However, it has been argued in Ref.~\cite{alba2021diffusion} that the growth of the $OSE$ entropy reflects the diffusive quasiparticle 
spreading. Indeed, as a consequence of the 
interactions, the trajectory of the solitons (both the ones produced at the center and the background ones)  
exhibits diffusion. If one imagine of coarse graining the dynamics, one has that the 
effect of interactions, at long times and large distances, is a renormalization of the soliton velocities~\cite{buca2021rule}. 
In the coarse-grained picture, the solitons move with their renormalized velocities as free particles, i.e., 
without scattering. Now, the dynamics resembles that of the identity operator, implying that 
the operator entanglement does not grow. This suggests that the growth of the $OSE$ 
could be attributed to the diffusive fluctuations of the solitons trajectories.

\subsection{Spin-$1/2$ $XXZ$ chain and its nonintegrable deformation}
\label{sec:xxz}

In this work we also consider a nonintegrable deformation of the spin-$1/2$ Heisenberg $XXZ$ 
chain defined by the Hamiltonian 
\begin{equation}
	\label{eq:xxz-ham}
	H=\sum_{x=1}^{L-1}\frac{1}{2}\left(S_x^+S_{x+1}^-+S_{x}^-S_{x+1}^+\right)
	+\Delta\sum_{x=1}^{L-1} S_{x}^zS_{x+1}^z+\Delta'\sum_{x=1}^{L-2}S_{x}^zS_{x+2}^z. 
\end{equation}
Here $S_x^{\pm}$ are standard raising/lowering spin-$1/2$ operators acting at site $x$, and $\Delta,\Delta'$ 
are real parameters. For $\Delta=\Delta'=0$ the model~\eqref{eq:xxz-ham} can be mapped to a system of 
noninteracting fermions. 
For $\Delta'=0$, Eq.~\eqref{eq:xxz-ham} becomes the $XXZ$ chain, which 
is a prototypical Bethe ansatz solvable model~\cite{takahashi1999thermodynamics}. For nonzero 
$\Delta'$ the model is nonintegrable, as one can show by studying the statistics of the 
gaps in the energy spectrum~\cite{alba2021diffusion}. Notice that Eq.~\eqref{eq:xxz-ham} commutes with 
the total magnetization $M=\sum_x S^z_x$, also for nonzero $\Delta'$. As for the rule $54$ chain, we focus on the 
spreading of the diagonal operators $S^z$ and $P_z$, defined in~\eqref{eq:Pz-Sz}. The $XXZ$ chain 
possesses well-defined quasiparticles, which are at the heart of the quasiparticle 
picture~\cite{alba2017entanglement,alba2018entanglement} for entanglement spreading and 
of Generalized Hydrodynamics~\cite{bertini-2016,olalla-2016}. In contrast with the rule $54$ chain, 
for which one can construct explicitly the $MPO$ describing a generic time-dependent operator, 
it is much more challenging to obtain exact results for operator spreading in the 
$XXZ$ chain~\eqref{eq:xxz-ham}, even in the integrable case for $\Delta'=0$. 
However, it has been numerically shown in Ref.~\cite{alba2019operator} that for $\Delta'=0$ 
the $OSE$ entropy grows logarithmically with time for generic $\Delta$. In particular, 
Ref.~\cite{alba2021diffusion} showed that for $\Delta'=0$, $S_1(P_z)$ grows as $1/2\ln(t)$, 
for any $\Delta$, which is the same behavior observed in the rule $54$ chain. The dynamics of 
$S_1(S^z)$ is compatible with a logarithmic growth as $S_1(S^z)\approx \ln(t)$. 
Finally, for $\Delta'\ne0$, numerical results for the growth of the $OSE$ are inconclusive 
due to strong finite-time effects~\cite{alba2021diffusion}.

\section{Analytical bounds on the growth of the $OSE$ entropies}  
\label{sec:bound}

Let us now discuss some analytical results that are available for the $OSE$ entropies in 
generic systems. In Section~\ref{sec:muth} we review the result of Ref.~\cite{muth2011dynamical}, 
showing that the $OSE$ entropies of a given operator ${\mathcal O}$ can be bound by 
its  Infinite Temperature Auto Correlator ($ITAC$). In Section~\ref{sec:traceful}, following 
Ref.~\cite{jonay2018coarse} we argue that the $OSE$ of an operator is determined by its 
traceless part. Finally, in Section~\ref{sec:diffusion} we review the results 
of Ref.~\cite{alba2021diffusion} trying to establish a relationship between the growth of 
$OSE$ and diffusion.

\subsection{Operator entanglement and $ITAC$: The bound of Ref.~\cite{muth2011dynamical}}
\label{sec:muth}

Ref.~\cite{muth2011dynamical} showed that for $n>1$ the R\'enyi $OSE$ entropies $S_n(\hat{\mathcal O})$ of 
a generic local operator $\hat{\mathcal O}$ is bound by its $ITAC$ as 
\begin{equation}
	\label{eq:ineq}
	S_n(\hat{\mathcal O})\le \frac{2n}{1-n}\ln\left|\langle \hat{\mathcal O}^\dagger(t)\hat{\mathcal O}(0)
	\rangle_{T=\infty}\right|=\frac{2n}{1-n}\ln\left|\mathrm{Tr}(\hat{\mathcal{O}}^\dagger (t)\hat{\mathcal{O}}(0))\right|,\quad n>1, 
\end{equation}
where we used that $\langle \hat{\mathcal O}_1\hat {\mathcal O}_2\rangle_{T=\infty}=\mathrm{Tr}
(\hat{\mathcal O}_1 \hat {\mathcal O}_2)$. 
The derivation of~\eqref{eq:ineq} is straightforward, and we report it for completeness. 
One starts from the decomposition 
\begin{equation}
	\frac{\hat{\mathcal O}(t)}{\sqrt{\mathrm{Tr}(\hat{\mathcal O}^\dagger\hat{\mathcal O})}}=\sum_{rs}\Lambda_{rs}(t)
	\hat {\mathcal O}_{r,A}\otimes \hat{\mathcal O}_{s,B}, 
\end{equation}
where $\Lambda_{rs}$ is a time-dependent matrix, and $\hat{\mathcal O}_{r,A}$ and $\hat{\mathcal O}_{s,B}$ form 
two time-independent orthonormal bases for the operators with support in $A$ and $B$, i.e., satisfying 
$\mathrm{Tr}(\hat {\mathcal O}_{r,A(B)}^\dagger \hat {\mathcal O}_{r',A(B)})=\delta_{rr'}$. One can perform 
a Singular Value Decomposition of $\Lambda_{rs}$ to obtain 
\begin{equation}
	\label{eq:schmidt-2}
	\frac{\hat{\mathcal O}(t)}{\sqrt{\mathrm{Tr}(\hat{\mathcal O}^\dagger\hat{\mathcal O})}}=
	\sum_{r}\sqrt{\lambda_r(t)}\hat{ \widetilde{\mathcal O}}_{r,A}(t)\otimes \hat{\widetilde{\mathcal O}}_{r,B}(t), 
\end{equation}
where $\sqrt{\lambda_r}$ are the Schmidt values, and the operators appearing in the right-hand side 
of~\eqref{eq:schmidt-2} depend on time and are orthonormal. 
	Notice that the singular values (Schmidt values) 
are in one-to-one correspondence with the eigenvalues of the reduced density matrix of one of the subystems, 
and hence they are independent on the basis used to define $\Lambda_{rs}$. 
We now obtain the chain of inequalities as~\cite{muth2011dynamical} 
\begin{multline}
	\label{eq:chain-ineq}
	\left|\langle \hat{\mathcal O}^\dagger(t)\hat{\mathcal O}(0)\rangle_{T=\infty}\right|=
	|\mathrm{Tr}(\Lambda^\dagger(t)\Lambda(0))|\le|\sum_k \sqrt{\lambda_k(t)\lambda_k(0)}| \\
	\le \kappa_0\left(\sum_k \frac{\sqrt{\lambda_k(0)}}{\kappa_0}
	\lambda_k^{n/2}\right)^{1/n}	\le \kappa_0^{1-1/n}
	\left(\sum_k\lambda_k^n\right)^{1/(2n)},\quad \kappa_0:=\sum_k\sqrt{\lambda_k(0)}. 
\end{multline}
In the first row in~\eqref{eq:chain-ineq} we used the von Neuman trace inequality~\cite{mirsky1975a} 
$|\mathrm{Tr}(XY)|\le \sum_j\alpha_j\beta_j$, with $\alpha_j,\beta_j$ the singular values obtained 
from the Schmidt decomposition of $X$ and $Y$, respectively. 
In the second row of~\eqref{eq:chain-ineq} 
we employed the Jensen inequality 
\begin{equation}
	\varphi\left(\frac{\sum_k a_k x_k}{\sum_k a_k}\right)\ge \frac{\sum_k a_k
	\varphi(x_k)}{\sum_k a_k}, \quad \forall\varphi(x)\,\,\mathrm{concave}, 
\end{equation}
where we used $\varphi(x)=x^{1/n}$, $x_k=\lambda_k^{n/2}$,  and $a_k=\sqrt{\lambda_k(0)}$. 
We also used that $x^{1/n}$ is concave for $n>1$. 
In the last step we used the Cauchy-Schwartz inequality
\begin{equation}
	\sqrt{\sum_k u_k v_k}\le \sqrt{\sum_k u_k^2}\sqrt{\sum_k v_k^2}, 
\end{equation}
with $u_k=\sqrt{\lambda_k(0)}/\kappa_0$, and $v_k=\lambda_k^{n/2}$. 
In deriving~\eqref{eq:chain-ineq} we also used $\sum_k\lambda_k(0)=1$. 
For local product operators 
the Schmidt decomposition at $t=0$ is trivial with only one 
Schmidt value $\lambda=1$, implying $\kappa_0=1$. By taking the logarithm of both members of~\eqref{eq:chain-ineq} and multiplying 
them by $2n/(1-n)$, one obtains~\eqref{eq:ineq}.

Let us now focus on the $ITAC$ on the 
left-hand side of~\eqref{eq:ineq}. One should expect a vanishing behavior  in the limit $t\to\infty$, 
although several scenarios are possible. Ballistic spreading of operators, as in integrable systems, 
suggests power-law~\cite{doyon2020lecture} decay as $1/t$ in the limit $t\to\infty$, whereas diffusive spreading corresponds 
to the slower decay as $1/\sqrt{t}$. From~\eqref{eq:ineq}, power-law decay of the $ITAC$ implies a logarithmic growth of 
the $OSE$. 
On the other hand, for some operators the $ITAC$ in~\eqref{eq:ineq} decays exponentially 
with time. For instance, this happens  for the correlator $\langle S^+(t)S^-\rangle_{T=\infty}$ in the 
$XX$ spin chain~\cite{delvecchio2022the}. In these situations, 
Eq.~\eqref{eq:ineq} gives  a linear bound for  the $OSE$ entropies. 
However, the bound~\eqref{eq:ineq} is generically not saturated. For instance, it is well-know that in the 
$XX$ chain the $OSE$ entropy of $S^+$ grows logarithmically with 
time~\cite{pizorn2009operator,prosen2007operator,dubail2017entanglement}. 
Similarly, by assuming that  Eq.~\eqref{eq:ineq} is saturated for $S^z$, one has a logarithmic growth of the $OSE$, as 
$S_n(S^z)\sim 2n/(n-1)\ln(t)$ since $\langle S^z(t)S^z(0)\rangle\sim 1/t$ at long times~\cite{delvecchio2022the}. 
On the other hand, it is well-known that $S_n(S^z)$ saturates to a constant
at long times~\cite{pizorn2009operator,dubail2017entanglement}. We conclude that Eq.~\eqref{eq:ineq} 
is in general a loose bound on the $OSE$ growth. 

\subsection{Traceless versus traceful operators (Ref.~\cite{jonay2018coarse})} 
\label{sec:traceful}

Following Ref.~\cite{jonay2018coarse}, one can argue that the dynamics  of 
$S_n(\hat{\mathcal O})$ can be obtained from $S_n(\hat{\mathcal O'})$, where $\hat{\mathcal O}'$ 
is the traceless part of $\hat {\mathcal O}$. 
Let us decompose the generic  normalized operator $\hat{\mathcal O}$ as 
\begin{equation}
	\label{eq:cdec}
	\hat{\mathcal{O}}:= \sqrt{p}\mathds{1}+\sqrt{1-p}
	\hat{\mathcal{O}}'\quad \sqrt{p}:=\mathrm{Tr}(\hat{\mathcal{O}}), 
\end{equation}
where $\hat {\mathcal{O}}'$ is traceless and both $\hat{\mathcal{O}}'$ and $\mathds{1}$ are 
normalized. 
The time-evolved operator $\hat {\mathcal O}(t)$ can be decomposed as 
\begin{equation}
	\label{eq:decomp}
	\hat{\mathcal{O}}(t):= \sqrt{p}\mathds{1}+\sqrt{1-p}
	\hat{\mathcal{O}}'(t), 
\end{equation}
and $\hat {\mathcal O}'(t)$ is traceless, since the dynamics preserves the trace. 
After performing a Schmidt decomposition of $\hat{\mathcal O}'$, we obtain 
\begin{equation}
	\label{eq:decomp-1}
	\hat{\mathcal O}(t)=
	\sqrt{p}\mathds{1}_A\otimes \mathds{1}_B+
	\sum_i\sqrt{(1-p)\lambda'}_i(t)\hat{\mathcal O}'_{A,i}(t)\otimes\hat{\mathcal O}'_{B,i}(t), 
\end{equation}
where $\lambda'_i(t)\ge 0$ are the Schmidt values. 
	As discussed in Ref.~\cite{jonay2018coarse} while at $t=0$ the Schmidt decomposition 
	in~\eqref{eq:cdec} is exact, the decomposition for $\hat{O}(t)$ is only 
	approximate. The reason is that the time-evolved 
	operators $\hat{\mathcal O}'_{i,A}(t)$ are not traceless. 

In general one should expect that at least some of the operators $\hat{\mathcal O}'_{A,i}$ 
appearing in the second term in~\eqref{eq:decomp-1}, and which are initially traceless, 
will develop a nonzero trace with time.  
This implies that they will not be orthogonal to the identity 
$\mathds{1}_A$, which makes it apparent that Eq.~\eqref{eq:decomp-1} is 
not a Schmidt decomposition for $\hat {\mathcal O}(t)$. 
If we assume that all the $\hat{\mathcal O}'_{A,i}$ are 
traceless, then Eq.~\eqref{eq:decomp-1} is a Schmidt decomposition 
for $\hat {\mathcal O}(t)$. Although this is an approximation,  it is natural to expect 
(see section~\ref{sec:trace-xxz}) that the probability that the dynamics 
generates the identity operator 
in all the sites of $A$ rapidly decays with the size of $A$, 
meaning that the decomposition~\eqref{eq:decomp-1} 
will become asymptotically exact in the limit $t\to\infty$ if $A$ is the 
semi-infinite system.

This implies that the operator entanglement spectrum $\{\lambda_i\}$ of $\hat{\mathcal O}(t)$ 
can be decomposed as 
\begin{equation}
	\label{eq:decomp}
	\{\lambda_j\}=\{p\}\cup \{(1-p)\lambda_i'\},
\end{equation}
where $\lambda_i$ are the same as in~\eqref{eq:decomp-1}. 
Eq.~\eqref{eq:vn-simp} implies that 
the $OSE$ of a given operator $\hat{\mathcal O}$ 
can be determined by that of its traceless part $\hat{\mathcal O}'$. 
Specifically, for the von Neumann entropy one has 
\begin{equation}
	\label{eq:vn-simp}
	S(\hat{\mathcal O})=p\ln(p)+(1-p)\sum_i\lambda'_i \ln(1-p)+(1-p)
	S(\hat{\mathcal O}')= p\ln(p)+(1-p)\ln(1-p)+(1-p)S(\hat{\mathcal O}'),  
\end{equation}
where in the last step we used that $\sum_i\lambda_i'=1$. 

Similarly, the R\'enyi entropy satisfy 
\begin{equation}
	\label{eq:renyi-simp}
	S_n(\hat {\mathcal O})=\frac{1}{1-n}\ln\Big[p^n+(1-p)\sum_i{\lambda'_i}^n\Big]\approx \frac{n}{1-n}\ln(p), 
\end{equation}
where we used that at long times $\sum_i{\lambda'}^n_i\to 0$ because the operator support increases, while 
$\sum_i\lambda'_i=1$ because the operators are normalized.  
As we will discuss in Section~\ref{sec:renyi-54} and Section~\ref{sec:xxz-num}, the 
decomposition~\eqref{eq:decomp} captures correctly the dynamics of the $OSE$, 
at least at long times and for large subsystems $A$.

\subsection{Operator space entanglement and diffusion (Ref.~\cite{alba2021diffusion})} 
\label{sec:diffusion}

Inspired by the results for the rule $54$ chain, a plausible scenario is that the dynamics of the operator 
entanglement reflects the diffusive dynamics of the underlying solitons~\cite{alba2021diffusion}, as discussed already 
in Section~\ref{sec:rule-54}. 
Again, at a ballistic level, or at a coarse-grained level,  
the left and right movers behave as noninteracting particles, moving with effective velocities, which are 
determined by the scatterings. This also applies to the solitons produced at the center of the 
chain by the operator insertion. 
Since at a ballistic level the solitons behave as free particles, the operator entanglement does 
not grow~\cite{medenjak2022operator}. Still, it is challenging to encapsulate this idea in a quantitative 
framework. As suggested in Ref.~\cite{alba2021diffusion}, 
a crude approximation for the Schmidt decomposition of an 
operator $\hat{\mathcal O}$ is 
\begin{equation}
	\label{eq:decomp-2}
	\hat {\mathcal O}(t)=\sum_{k=0}^{t-|x|}\frac{\sqrt{\binom{t-|x|}{k}
	\binom{t+|x|}{t-k}}}{\sqrt{\binom{2t}{t}}}
	\hat {\mathcal O}_{A,k}\otimes\hat{\mathcal O}_{B,t-k}.
\end{equation}
In~\eqref{eq:decomp-2}, $\hat{\mathcal O}_{A,k}$ and $\hat{\mathcal O}_{B,t-k}$ are 
normalized operators in $A$ and $B$ constructed with $k$ and $t-k$ solitons. In~\eqref{eq:decomp-2} we 
assume that $\hat{\mathcal O}_{A,k}$ and $\hat{\mathcal O}_{B,t-k}$ are some
``flat'' superpositions of all the configurations with $k$ and $t-k$ solitons. Physically, 
this means that the scatterings maximally ``scramble'' the trajectories of 
the solitons within the lightcone. 
In~\eqref{eq:decomp-2}, $x$ is the position of the cut dividing subsystem $A$ from $B$ (see Fig.~\ref{fig0:cartoon}). 
The two binomials in~\eqref{eq:decomp} are the number of ways of
arranging the  solitons in $A$ and $B$. In the limit $t\to\infty$ the coefficients in~\eqref{eq:decomp-2} 
are dominated by the term with $k=(t-|x|)/2$, with diffusive fluctuations as $\sqrt{t}$. 
Now, the decomposition in~\eqref{eq:decomp-2} is an approximation because 
interactions are local and the trajectories of the solitons cannot be 
scattered at arbitrary distances. 
The bond dimension of the decomposition in~\eqref{eq:decomp} is $t-|x|+1$. Notice that $t-|x|$ is the largest number of
solitons that can be accommodated within $A$. Notice also that Eq.~\eqref{eq:decomp-2} implies 
that there are only $\propto t$ nonzero Schmidt values, whereas the exact $MPO$ describing 
the dynamics of generic operators has $\sim t^2$ nonzero Schmidt values~\cite{alba2019operator}. 
By using~\eqref{eq:decomp-2}, one obtains the analytical result for the  von Neumann entropy  
as 
\begin{equation}
	\label{eq:bound}
	S_{n}= \frac{1}{2}\ln(t),\quad \forall n.
\end{equation}
Thus, Eq.~\eqref{eq:decomp-2} implies that the R\'enyi $OSE$ entropies should not depend on 
$n$. The prefactor 
$1/2$ of the logarithmic growth clearly reflects the diffusive spreading of the solitons trajectories. 
As it will be clear in the following, $S_n$ depend on $n$, signaling that 
Eq.~\eqref{eq:decomp-2} does not capture correctly the detailed structure of the operator entanglement 
spectrum. On the other hand, Eq.~\eqref{eq:bound} is in agreement with the growth of the 
von Neumann entropy of the projector operator $P_z$ (see Ref.~\cite{alba2021diffusion}).  

\section{R\'enyi entanglement of operators in the rule $54$ chain: A $tDMRG$ analysis}
\label{sec:renyi-54}

Here we discuss $tDMRG$ results for $S_n$ of diagonal operators in the rule $54$ chain. 
In particular, in Section~\ref{sec:trace} we show that given an operator $\hat{\mathcal O}$ 
with nonzero trace, $S_1(\hat{\mathcal O})$ can be obtained from $S_1(\hat{\mathcal O}')$, 
with $\hat{\mathcal O}'$ the traceless part of the operator. Moreover, we show that 
$S_n$ saturate to a constant value in the limit $t\to\infty$, in agreement with the results 
of Section~\ref{sec:traceful}. In Section~\ref{sec:noncft} we focus 
on traceless operators. We numerically show that $S_n$ is compatible with a logarithmic growth 
for any $n$, although the prefactor of the logarithmic growth depends on $n$ in a nontrivial manner. 
Finally, in Section~\ref{sec:spetrum-54} we discuss how this is reflected in the distribution of the 
levels of the operator entanglement spectrum.

\subsection{Dynamics of the R\'enyi entropies for $P_z$: Effect of a nonzero trace} 
\label{sec:trace}

%
\begin{figure}[t]
\begin{center}
\begin{minipage}{.45\textwidth}
\includegraphics[width=1\textwidth]{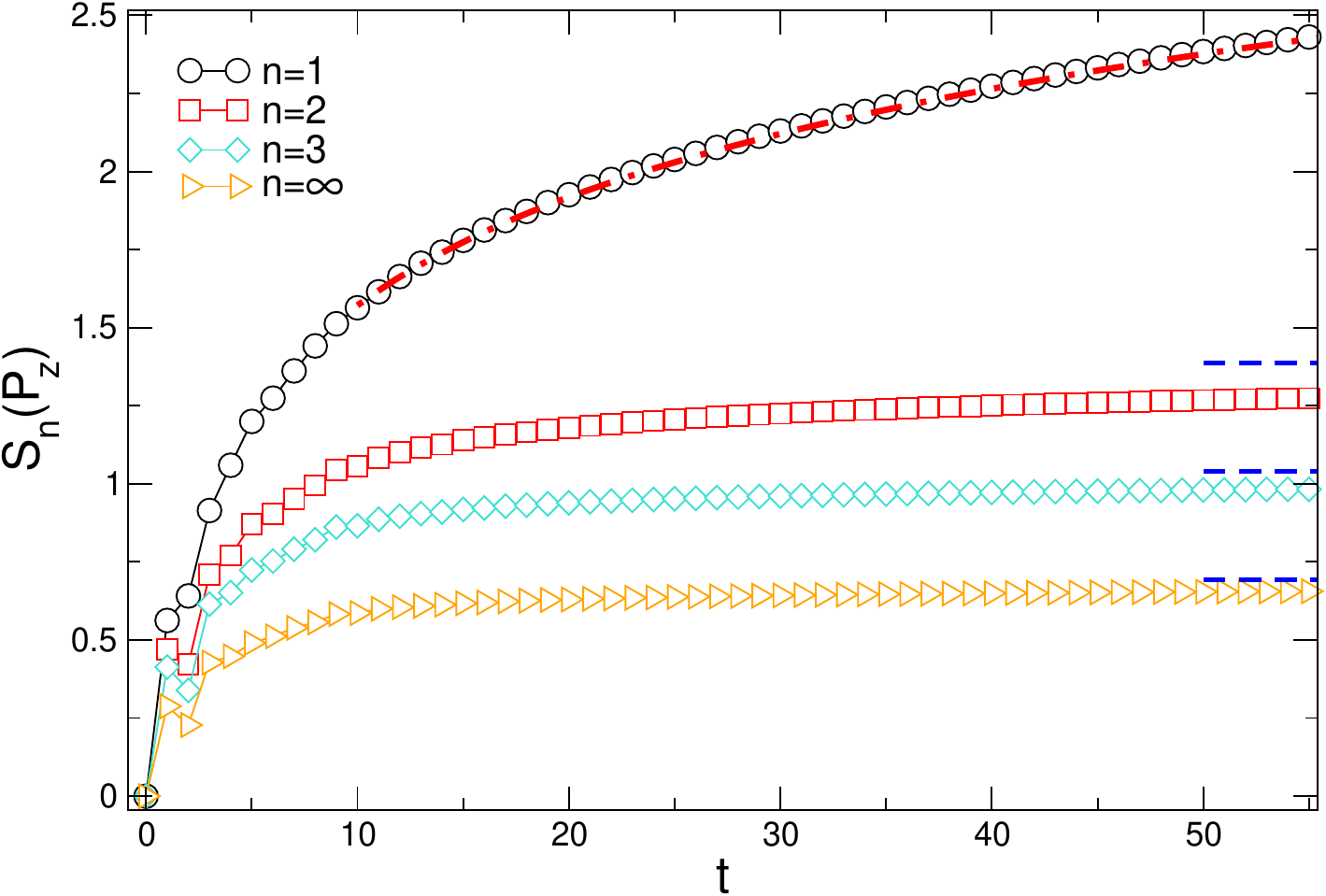}
\end{minipage}
\begin{minipage}{.45\textwidth}
\includegraphics[width=1\textwidth]{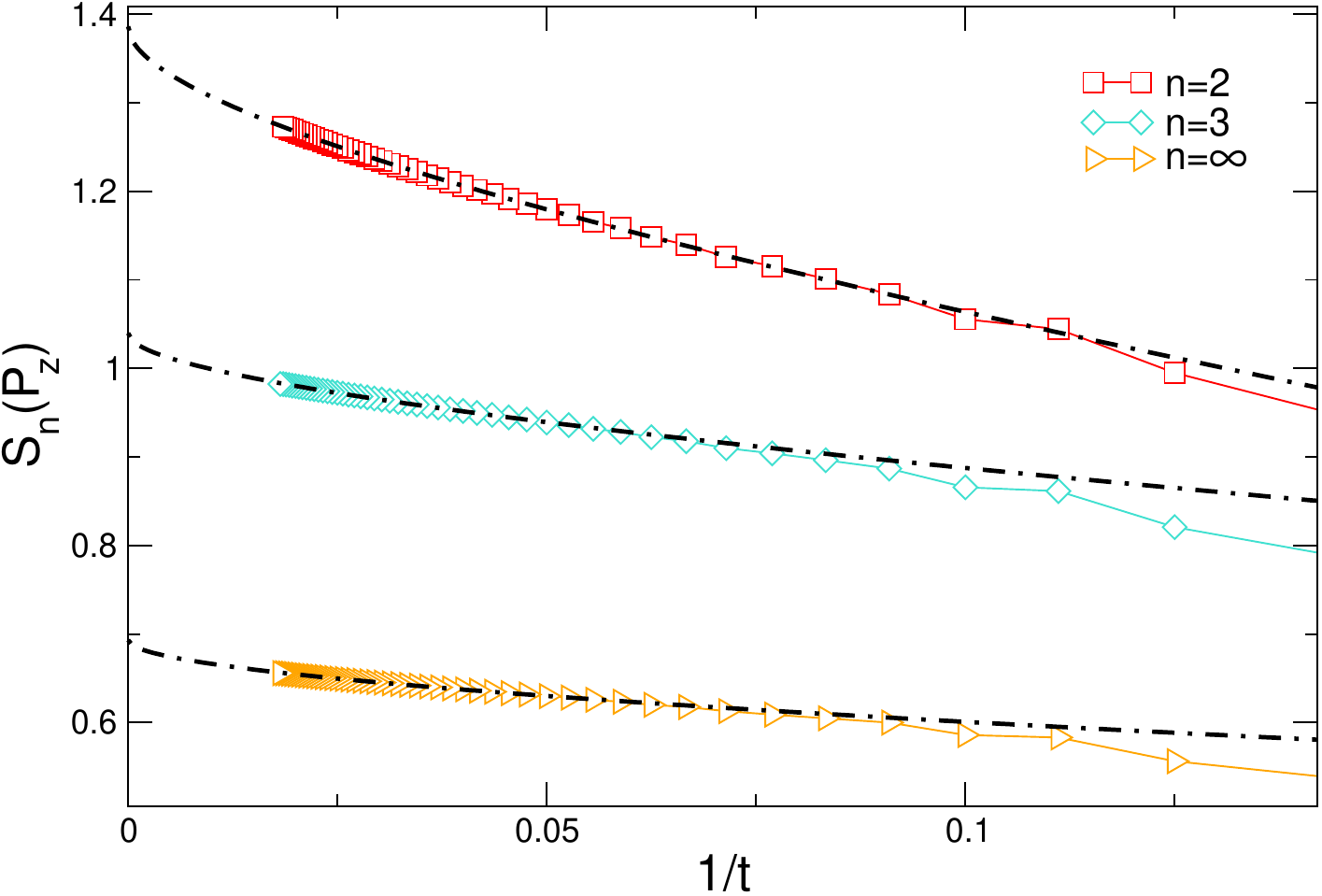}
\end{minipage}
\caption{Dynamics of the R\'enyi operator entanglement entropies $S_n$ for the projector operator $P_z$ in the rule $54$ chain. 
	Left panel:  $tDMRG$ data for $S_n$ plotted versus time $t$ for several values of $n$.  The dashed-dotted line is 
	the behavior $S_1=a\ln(t)+b$, with $a,b$ a fitting parameter. The fit gives $a=0.50(1)$. 
	The horizontal lines are $S_n=n/(n-1)\ln(p)$ 
	(cf.~Eq.~\eqref{eq:renyi-simp}) with $p=1/2$. Right panel: Same data 
	as in the left panel plotting $S_n-n/(n-1)\ln(2)$ versus $1/t$ for $n>1$. The dashed-dotted lines are fits 
	to $S_n=n/(n-1)\ln(2)+a_1/t^{1/2}+a_2/t$, with $a_1,a_2$ fitting parameters. 
}
\label{fig1:bobenko-tr}
\end{center}
\end{figure}
%
In Fig.~\ref{fig1:bobenko-tr} we show $tDMRG$ data for $S_n$ of the spin projector operator $P_z$ (cf.~\eqref{eq:Pz-Sz}) 
Our $tDMRG$ simulations are performed at fixed bond dimension $\chi=1600$, which allows us to 
reach times $t\approx 60$. As it is clear from Fig.~\ref{fig1:bobenko-tr} (left panel), 
$S_n(P_z)$ increase with time, although the dynamics is much slower for $n>1$, 
as compared with $n=1$. The growth 
of $S_1(P_z)$ was investigated  in Ref.~\cite{alba2019operator} and Ref.~\cite{alba2021diffusion}, and 
it is compatible with the behavior $S_1(P_z)\sim 1/2\ln(t)$, i.e., Eq.~\eqref{eq:bound}. 
The dashed-dotted line in the left panel of Fig.~\ref{fig1:bobenko-tr}  is 
\begin{equation}
	\label{eq:fit}
	S_n=a\ln(t)+b, 
\end{equation}
with $a,b$  fitting parameters. 
To obtain $a,b$ we fitted the data in the range $t_{min}\le t\le 50$ to~\eqref{eq:fit}, i.e., 
discarding the short times with $t\le t_{min}$ because Eq.~\eqref{eq:fit} is valid only at asymptotically 
long times. We estimated the error on $a$ by monitoring its variation  when increasing $t_{min}$. 
We should observe that by fitting the data for $n>1$ to~\eqref{eq:fit} we obtain  
quite ``small'' values for $a$ as $a\approx 0.08,0.04,0.02$ for $n,2,3,\infty$. However, as we now discuss, 
this signals that Eq.~\eqref{eq:fit} does not apply for $n>1$, and $S_n$ saturates for $t\to\infty$, 
as discussed in section~\ref{sec:traceful}. The saturation values predicted by~\eqref{eq:renyi-simp} 
are reported as horizontal dashed lines in Fig.~\ref{fig1:bobenko-tr}. The deviations from~\eqref{eq:renyi-simp} 
have to be interpreted as finite-time corrections. Corrections are more precisely investigated in 
the right panel of Fig.~\ref{fig1:bobenko-tr} plotting 
$S_n-n/(n-1)\ln(2)$ versus $1/t$. 
A fit of the data at long times to the behavior $\sim t^{-\alpha}$ gives $\alpha\approx 1/2$. 
This motivates the ansatz for $S_n$ as 
\begin{equation}
	\label{eq:fit-2}
	S_n=\frac{n}{n-1}\ln(2)+\frac{a_1}{t^{1/2}}+\frac{a_2}{t^{}}, 
\end{equation}
with $a_1,a_2$ fitting parameters. As it is clear from Fig.~\ref{fig1:bobenko-tr}, 
the fits are in satisfactory agreement with the numerical 
data, confirming that $S_n$ saturate to a constant in the limit $t\to\infty$. We anticipate that 
the saturating behavior of $S_n$ for $n>1$ is also confirmed by the behavior  of the operator 
entanglement spectrum (see Section~\ref{sec:spetrum-54}).

%
\begin{figure}[t]
\begin{center}
\includegraphics[width=.45\textwidth]{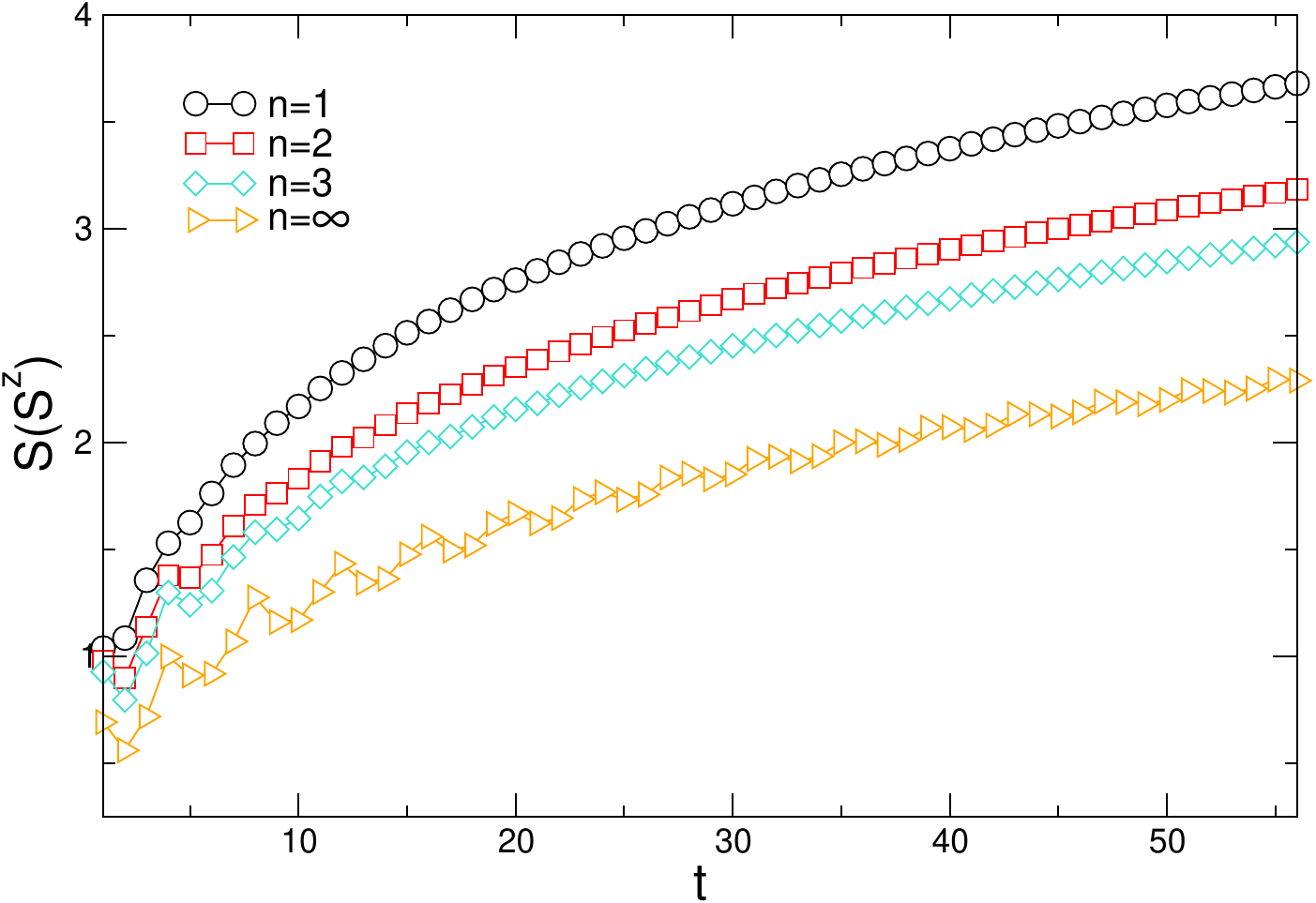}
\includegraphics[width=.45\textwidth]{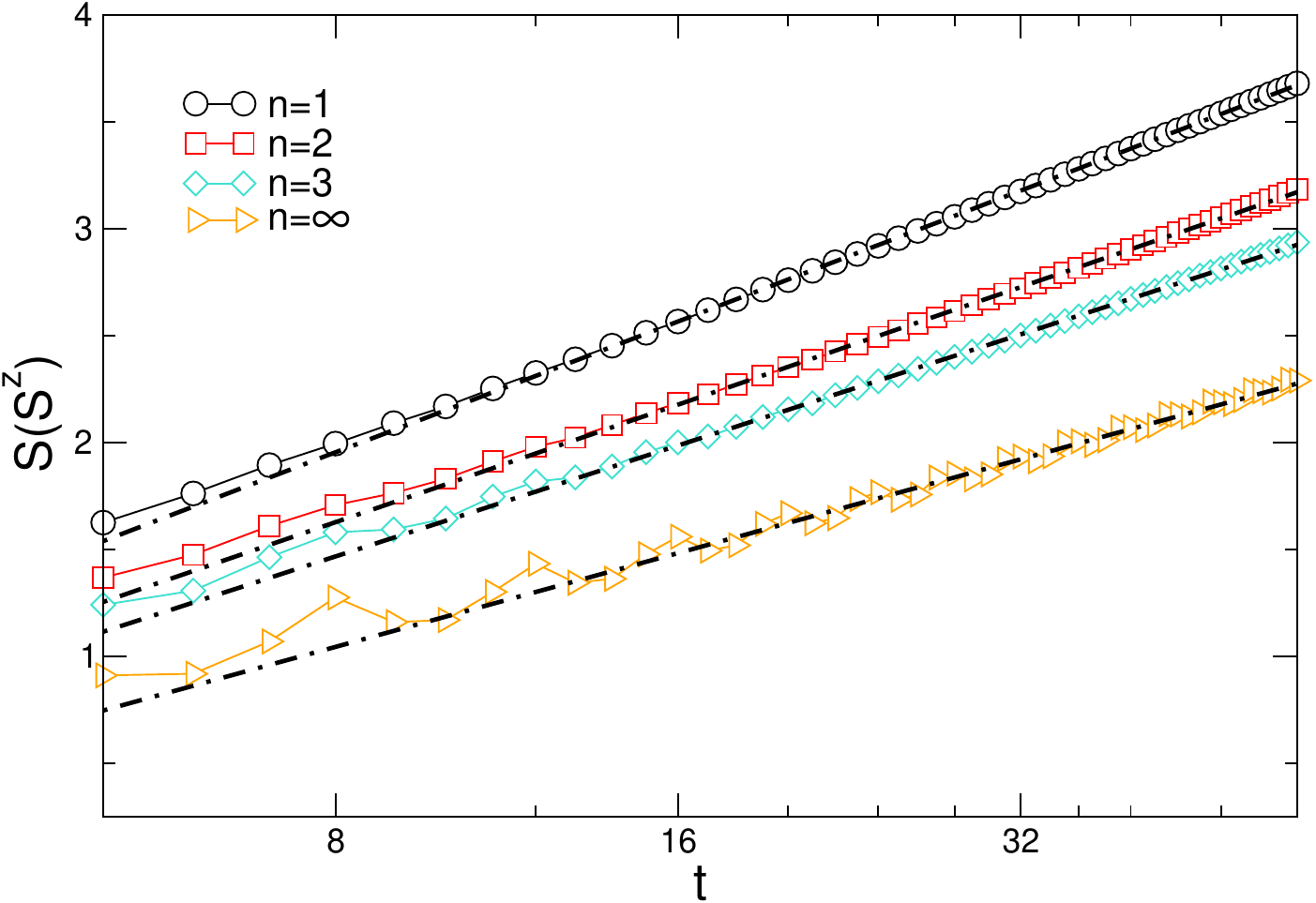}
\caption{Dynamics of the R\'enyi operator entanglement entropies $S_n$ for  $S^z$ in the rule $54$ chain. Left panel: 
	$tDMRG$ data for $S_n$ plotted versus time $t$ and $n=1,2,3,\infty$. Right panel: 
	Same as in the left panel using the logarithmic scale on the $x$-axis. 
	The dash-dotted lines are fits to $S_n=a\ln(t)+b$, 
	with $a,b$ fitting parameters. 
}
\label{fig2:bobenko}
\end{center}
\end{figure}
%

\subsection{Non $CFT$ scaling of the $OSE$ of traceless operators?}
\label{sec:noncft}

Having discussed the R\'enyi entropies of $P_z$, we now focus on its traceless part $S^z$. 
In Fig.~\ref{fig2:bobenko} we show $tDMRG$ data for the rule $54$ chain. Our data are 
obtained by employing bond dimension $\chi=1600$. In contrast with the projector operator $P_z$ 
(see Fig.~\ref{fig1:bobenko-tr}), now $S_n$ grow logarithmically also for $n>1$. This is confirmed in 
the right panel of Fig.~\ref{fig2:bobenko} plotting $S_n$ versus time by using a logarithmic scale 
on the $x$-axis. 
The dashed-dotted lines in Fig.~\ref{fig2:bobenko} are fits to the behavior~\eqref{eq:fit}. 
We now obtain $a=0.92(2),0.80(5),0.76(5),0.64(5)$ for $n=1,2,3,\infty$. 
Notice that for $n=1$ the fit gives $a=0.92(2)$, which appears to be different from $a=1$. 
This is also in contrast with what one should expect from~\eqref{eq:vn-simp}, which would give $a\approx 1$ 
since for $P_z$ one has $a\approx 0.5$.  Still, the discrepancy could be attributed to the 
systematic error due to finite-time corrections. 

\begin{figure}[t]
\begin{center}
\includegraphics[width=.55\textwidth]{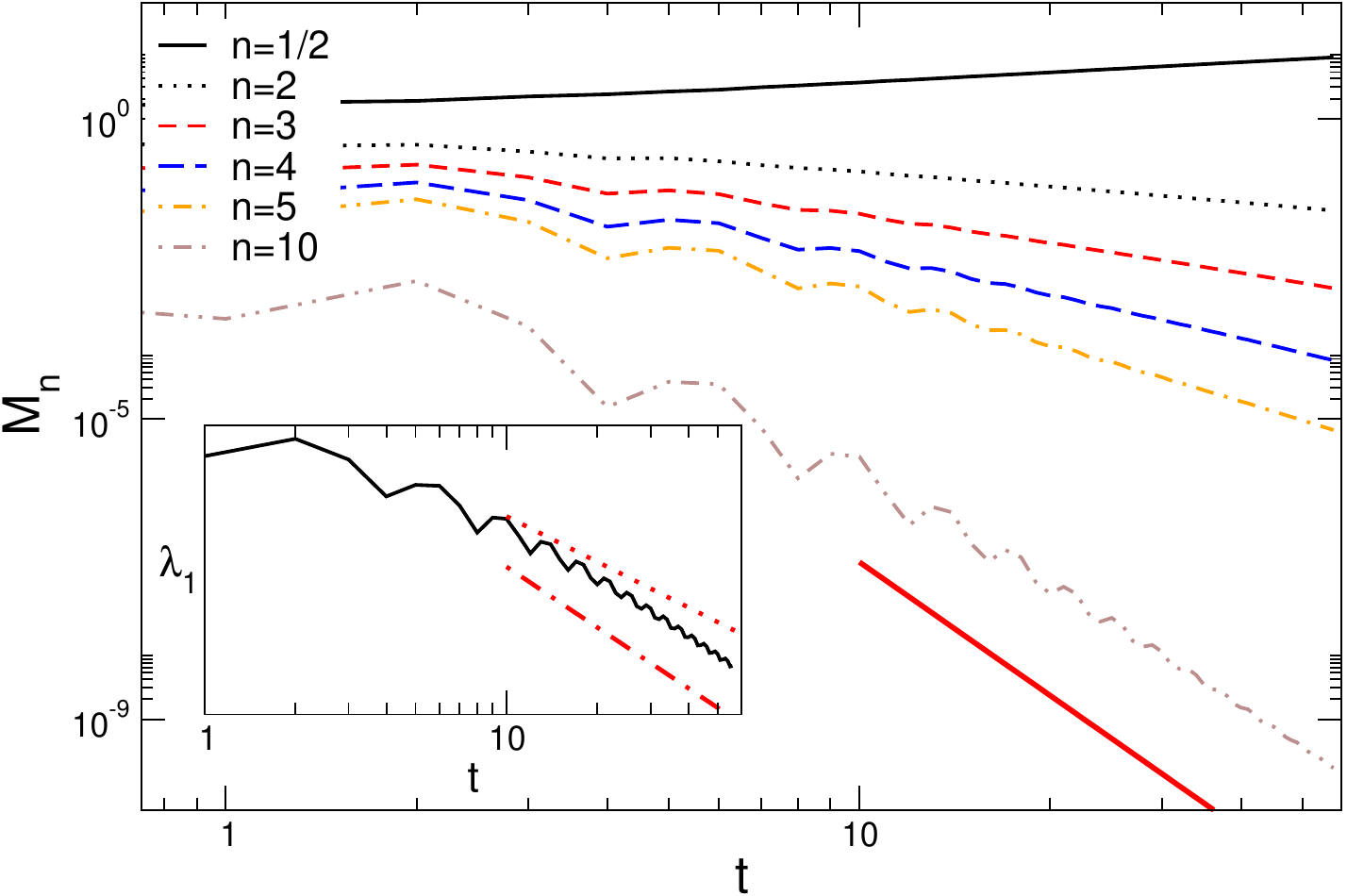}
\caption{Moments $M_n$ of the squared singular values $\lambda_i$ (cf.~\eqref{eq:schmidt}) for the 
	dynamics of the operator $S^z$ in the rule $54$ chain. We plot $M_n$ as functions of time $t$ and 
	for several values of $n$. Notice the logarithmic scale on both axes. The thick red line is the 
	behavior $t^{-2/3 n}$ for $n=10$. The inset shows the largest Schmidt value $\lambda_1$ as a 
	function of $t$. The dotted line and the dashed-dotted lines are the behaviors $t^{-1/2}$ and 
	$t^{-2/3}$, respectively. 
}
\label{fig2-b:bobenko}
\end{center}
\end{figure}

Let us consider the scaling of the moments $M_n$ of the 
$\lambda_i$ (cf.~\eqref{eq:schmidt}), which form the operator entanglement spectrum. 
$M_n$ are defined as 
\begin{equation}
	M_n:=\sum_i \lambda_i^n. 
\end{equation}
Clearly, $M_n$ contain the same information about the distribution of the ES levels as the 
R\'enyi entropies. 
The moments $M_n$ for the operator $S_z$ are plotted in Fig.~\ref{fig2-b:bobenko}. In the Figure we 
show $M_n$ versus time for several values of $n$, employing a logarithmic scale on both axis. First, the data exhibit 
a clear power-law behavior with increasing $t$. Moreover, for $n<1$, $M_n$ increases with time, whereas for $n>1$, 
$M_n$ vanish in the limit $t\to\infty$. The exponent of the power-law depends on $n$, reflecting that 
the prefactor of the logarithmic growth of $S_n$ depends on $n$ (see Fig.~\ref{fig2-b:bobenko}). For $n=1$, one has the normalization 
$M_1=1$ (not shown in the Figure). In the limit $n\to\infty$ a fit to the power-law behavior 
suggests the decay as $t^{-2/3n}$. The large $n$ limit is confirmed in the inset of 
Fig.~\ref{fig2-b:bobenko}, where we plot $\lambda_1$ versus $t$.  The dotted and the dashed-dotted 
lines are the behaviors as $t^{-1/2}$ and $t^{-2/3}$, respectively. 

Let us assume that $M_n$ for large $t$ have the power-law behavior as 
\begin{equation}
	\label{eq:Mn}
	M_n\sim t^{-f_n}.  
\end{equation}
It is natural to conjecture that $f_n$ is of the form 
\begin{equation}
	\label{eq:hyp}
	f_n=n a_0+\frac{a_1}{n}+\frac{a_2}{n^2}+\cdots, 
\end{equation}
where the dots are for higher powers of $1/n$. Eq.~\eqref{eq:hyp} is reminiscent of the $CFT$ scaling  
of the moments of the ground-state ES spectrum~\cite{calabrese2008entanglement}. Precisely, in $CFT$s one has 
that only $a_0$ and $a_1$ are nonzero, and $a_1=-a_0$, with $a_0$ related to the central charge of the 
$CFT$. Let us now consider the limit $n\to1$ to recover the von Neumann $OSE$. 
If we assume that only $a_0,a_1$ are nonzero in~\eqref{eq:hyp}, one has 
$S_1\sim 2a_0\ln(t)$. Fig.~\ref{fig2:bobenko} suggests the behavior $S_1\sim\ln(t)$, whereas 
from Fig.~\ref{fig2-b:bobenko} we have $\lambda_1\sim t^{-2/3}$, implying $a_0>1/2$. This 
discrepancy could signal that the terms $a_j$ with $j>1$ in the expansion~\eqref{eq:hyp} are nonzero. 
Finally, we should stress that $CFT$-like scaling was demonstrated 
for the growth of R\'enyi  entropies of string operators in the $XX$ chain~\cite{dubail2017entanglement}. 
%
%

\subsection{Operator entanglement spectrum ($ES$)}
\label{sec:spetrum-54}

\begin{figure}[t]
\begin{center}
\includegraphics[width=.45\textwidth]{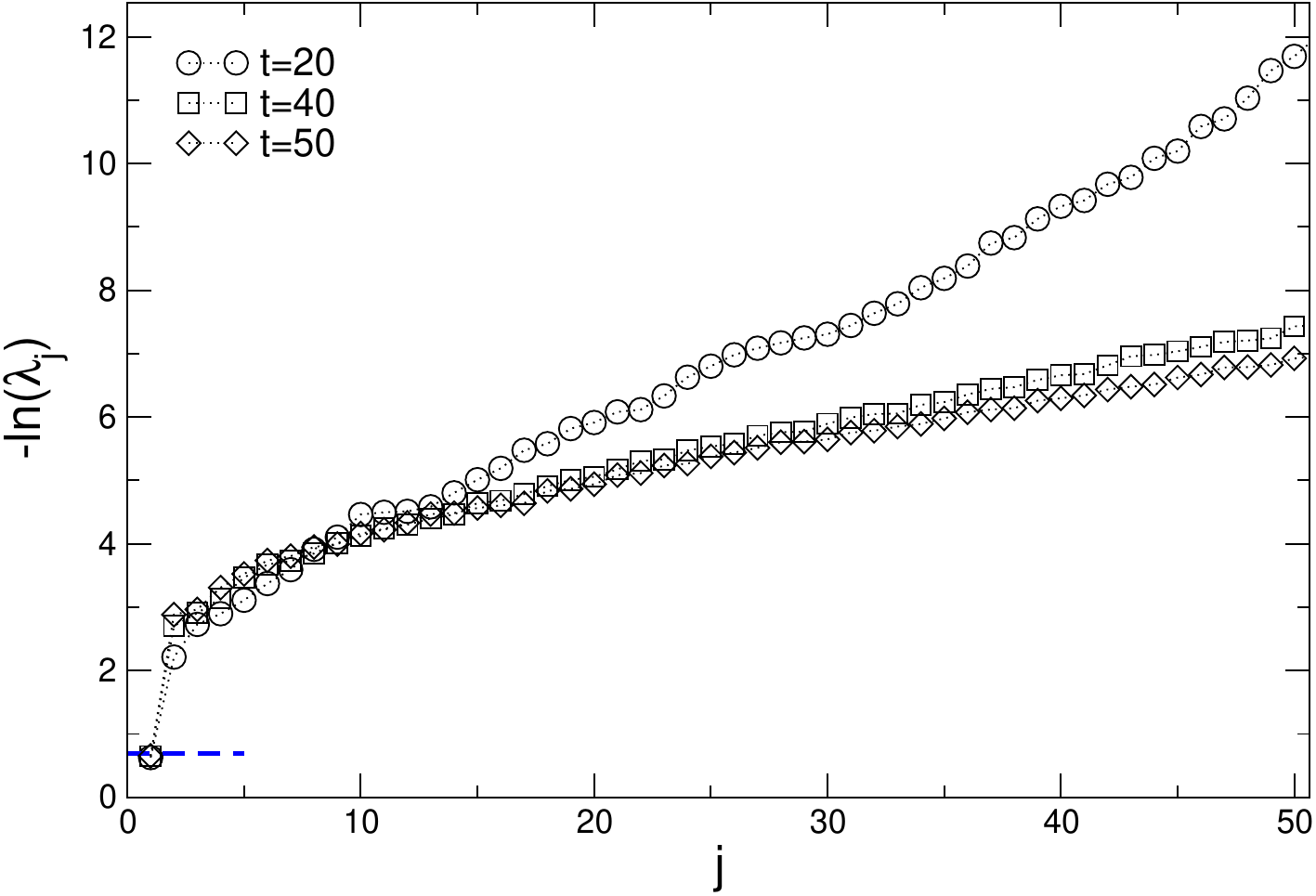}
\includegraphics[width=.45\textwidth]{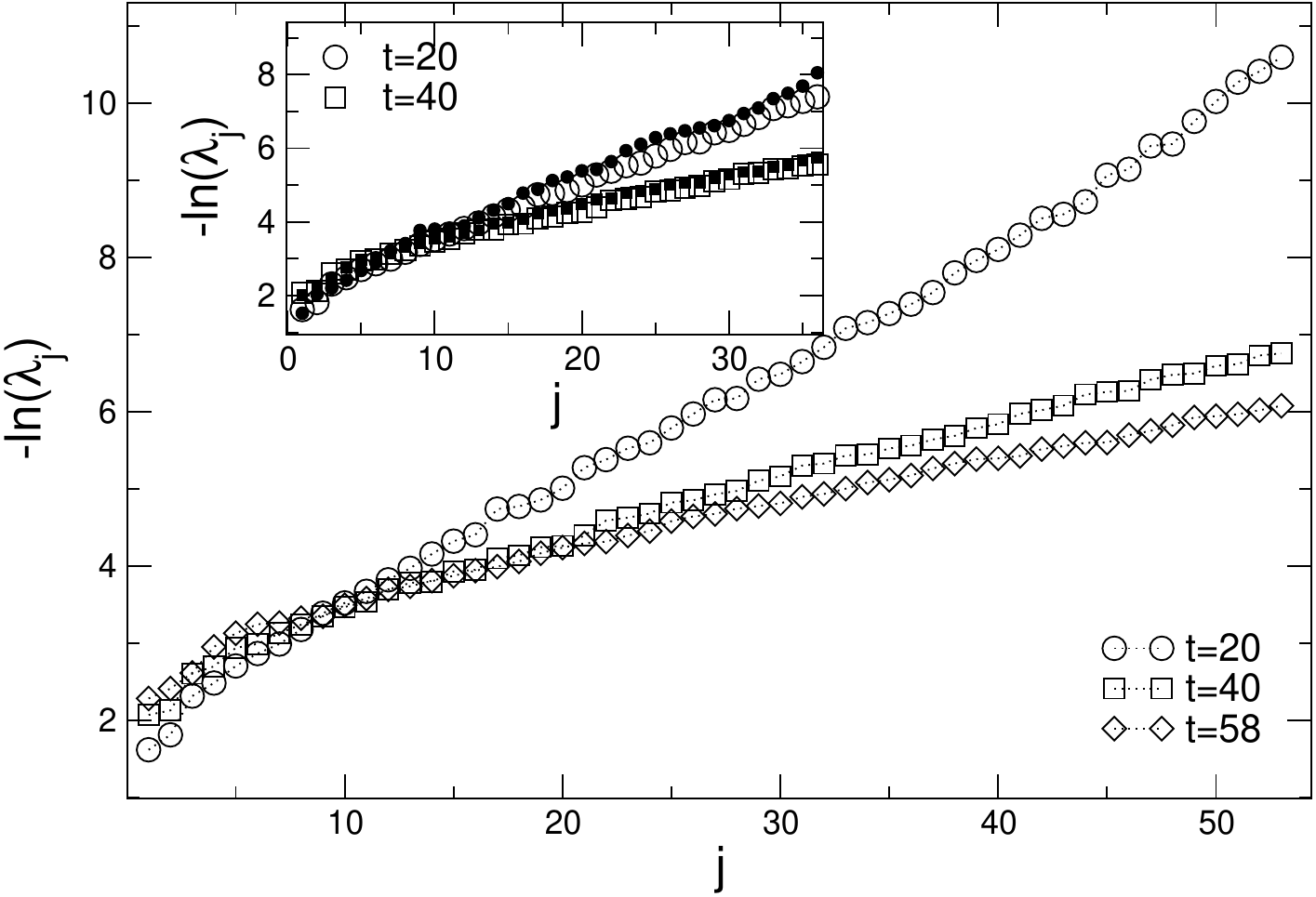}
\caption{ Dynamics of the operator entanglement spectrum in the rule $54$ chain. Both panels show $tDMRG$ results 
	with bond dimension $\chi=1600$. Left panel: Operator 
	entanglement spectrum for the projector operator $P_z$. The $y$-axis shows $-\ln(\lambda_j)$, 
	with $\lambda_j$ the squared Schmidt values. On the $x$-axis 
	$j$ is a label for the entanglement spectrum levels. The different symbols are for different 
	times. Right panel: Same as in the left panel for the operator $S^z$. Inset: Comparison between 
	the $ES$ levels $-\ln(\lambda'_j)$ of $S^z$ (empty symbols) and the shifted $ES$ 
	levels $-\ln(\lambda_j)-\ln(2)$ of $P_z$. 
}
\label{fig7:spect}
\end{center}
\end{figure}

Let us  now consider the operator entanglement spectrum ($ES$). The levels $\xi_i$ of the  $ES$ are defined as 
\begin{equation}
	\xi_i=-\ln(\lambda_i),
\end{equation}
with $\lambda_i$ the squared singular values in~\eqref{eq:schmidt}. In Fig.~\ref{fig7:spect} we 
focus on the dynamics of the entanglement spectrum of $P_z$ and of $S^z$ (left and right panels, respectively). 
Notice that $P_z$ and $S^z$ differ by $1/2\mathds{1}$, and the results of Section~\ref{sec:traceful} should hold. 
The $ES$ levels exhibit a 
nontrivial dynamics. Upon increasing time, there are more and more levels that contribute  to 
the $ES$. 
Let us focus on the operator entanglement spectrum of $P_z$ (left panel in Fig.~\ref{fig7:spect}). The smallest $ES$ level is 
$-\ln(\lambda_1)\approx\ln(2)$, reflecting the behavior  
$S_\infty\to \ln(2)$ for $t\to\infty$, which we observed in Fig.~\ref{fig2:bobenko} (right panel). 
The $ES$ levels above the lowest one are strongly suppressed. For instance, for $t=40$ they are 
$-\ln(\lambda_2)\approx 2.7$.  This is in agreement with the decomposition~\eqref{eq:decomp}. 

Moreover, according to~\eqref{eq:decomp} 
one can obtain the operator $ES$ of $P_z$ from that of $S^z$. 
Indeed, one should expect that (cf.~\eqref{eq:decomp})
\begin{equation}
	\lambda_{i}=\left\{\frac{1}{2},\frac{1}{2}\lambda_i'\right\}, 
\end{equation}
with $-\ln(\lambda'_i)$ forming the $ES$ of $S^z$ and $-\ln(\lambda_i)$ that of $P_z$. The operator $ES$ of $S^z$ is reported 
in the right panel in Fig.~\ref{fig7:spect}. Now the $ES$ level $\lambda_1=\ln(2)$ is not present, in contrast 
with the $ES$ of $P_z$ (left panel in the Figure). 
In the inset  we show the $ES$ levels for $S^z$ for $t=20,40$ (empty symbols). 
These are the same data reported in the main Figure. We also show the shifted levels $-\ln(\lambda_i)-\ln(2)$ 
with $\lambda_j$ forming the  $ES$ of $P_z$. Interestingly, they match the $ES$ of $S^z$ quite accurately, 
and the agreement improves at longer times.  
As anticipated, the results of Fig.~\ref{fig7:spect} confirm 
that although the decomposition in~\eqref{eq:decomp} is an approximation because 
$\hat{\mathcal{O}}'_{i,A}(t)$ (cf.~\eqref{eq:decomp-1}) have nonzero trace at $t>0$, 
it becomes exact in the limit $t\to\infty$. This is due to the fact that it is ``unlikely'' 
that the dynamics gives rise to a string of identity operators in the full subsystem $A$.

%
%
%
\begin{figure}[t]
\begin{center}
\includegraphics[width=.55\textwidth]{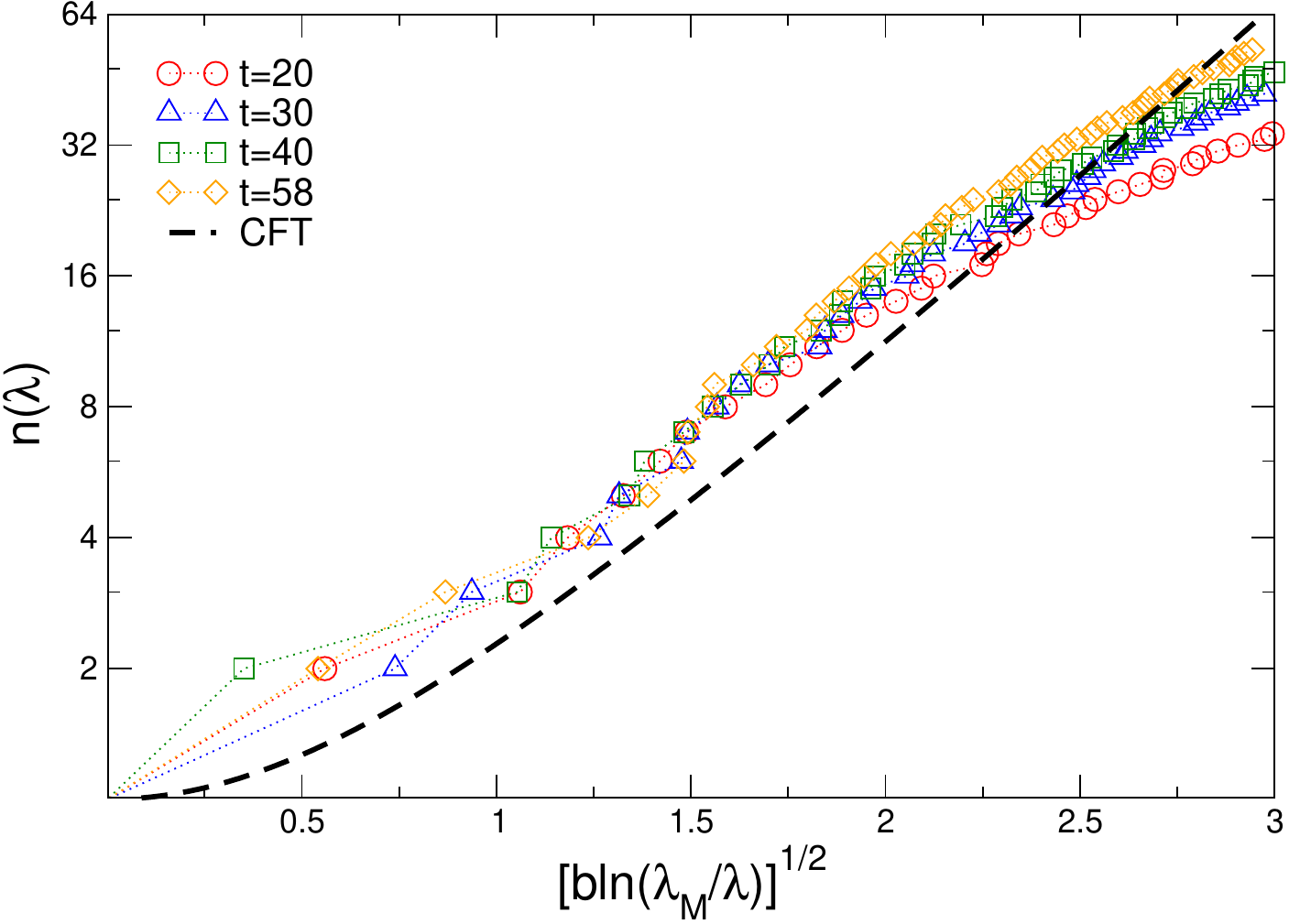}
\caption{Dynamics of the operator entanglement spectrum levels in the rule $54$ chain. 
	We show $tDMRG$ data at fixed bond dimension $\chi=1600$ for the operator $S^z$. 
	We plot $n(\lambda)$, i.e., the number of $ES$ levels smaller than 
	$\lambda$, as a function of the scaling variable $\xi=\sqrt{b\ln(\lambda_M/\lambda)}$, 
	with $b:=-\ln(\lambda_M)$, and $-\ln(\lambda_M)$ the lowest $ES$ level. 
	Different symbols correspond to different times. The dashed line is~\eqref{eq:nl}, i.e., 
	the CFT-like scaling. 
}
\label{fig8:spect_n}
\end{center}
\end{figure}
%
Finally, it is interesting to discuss the distribution of $ES$ levels. Since the $ES$ of 
operators with nonzero trace can be derived from that of their traceless part,  we focus on $S^z$. 
We consider the number of $ES$ levels $n(\lambda)$, which are larger than $\lambda$. In 
critical ground states described by $CFT$s, it has been shown that~\cite{calabrese2008entanglement} 
\begin{equation}
	\label{eq:nl}
	n(\lambda)=I_0(2\xi), \quad \xi:=\sqrt{b\ln(\lambda_M/\lambda)}, 
\end{equation}
where $\lambda_M$ is the largest eigenvalue of the $ES$, $b=-\ln(\lambda_M)$, and 
$I_0$ is the modified Bessel function of the first kind. 
Eq.~\eqref{eq:nl} implies that $n(\lambda)$ is ``superuniversal'', because it depends only 
on the central charge of the $CFT$,  which is encoded in the scaling of 
$\lambda_M$. 

In Fig.~\ref{fig8:spect_n} we plot $n(\lambda)$ as a function of $\xi$. The different symbols 
correspond to different times. At long times the data exhibit scaling collapse. This could suggest that 
although the moments $M_n$ do not scale as in $CFT$ (see Fig.~\ref{fig8:spect_n}), $n(\lambda)$ 
is a function of $\xi$, at least in the 
limit $t\to\infty$. The dashed line in Fig.~\ref{fig8:spect_n} is the prediction in $CFT$ systems 
$n(\lambda)=I_0(2\xi)$ (cf.~\eqref{eq:nl}). Fig.~\ref{fig8:spect_n} shows strong deviations from~\eqref{eq:nl}. 
This reflects that the moments $M_n$ (cf.~\eqref{eq:Mn}) do not obey $CFT$ scaling, which is given by  
Eq.~\eqref{eq:hyp} with $a_j=0$ for $j>1$ and $a_1=-a_0$.

\section{R\'enyi $OSE$ entropies in the deformed $XXZ$ chain: $tDMRG$ results}
\label{sec:xxz-num}

We now focus on the operator spreading in the deformed $XXZ$ chain (cf.~\eqref{eq:xxz-ham}). 
We first consider the integrable case with $\Delta'=0$, i.e., the $XXZ$ chain. 
In section~\ref{sec:trace-xxz} we show that the decomposition~\eqref{eq:decomp} holds true also 
for the $XXZ$ chain, similar to the rule $54$ chain (see section~\ref{sec:trace}). In section~\ref{sec:int-break} 
we investigate the effect of $\Delta'$, which breaks the integrability. Finally, in section~\ref{sec:revivals} 
we investigate revivals in the $OSE$ entropies in finite-size systems, showing that they can 
distinguish integrable from non-integrable dynamics. 

%
\begin{figure}[t]
\begin{center}
\includegraphics[width=.45\textwidth]{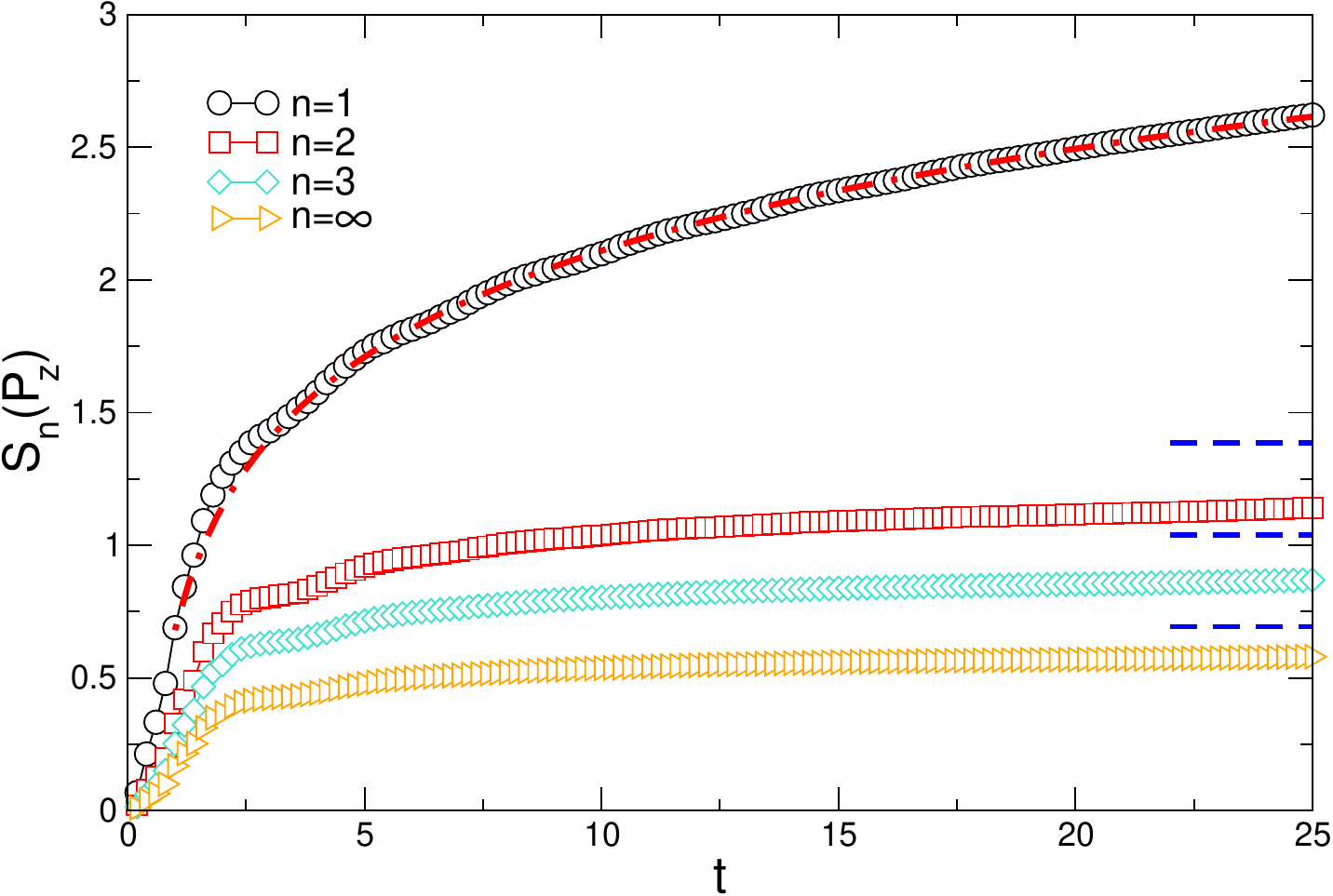}
\includegraphics[width=.45\textwidth]{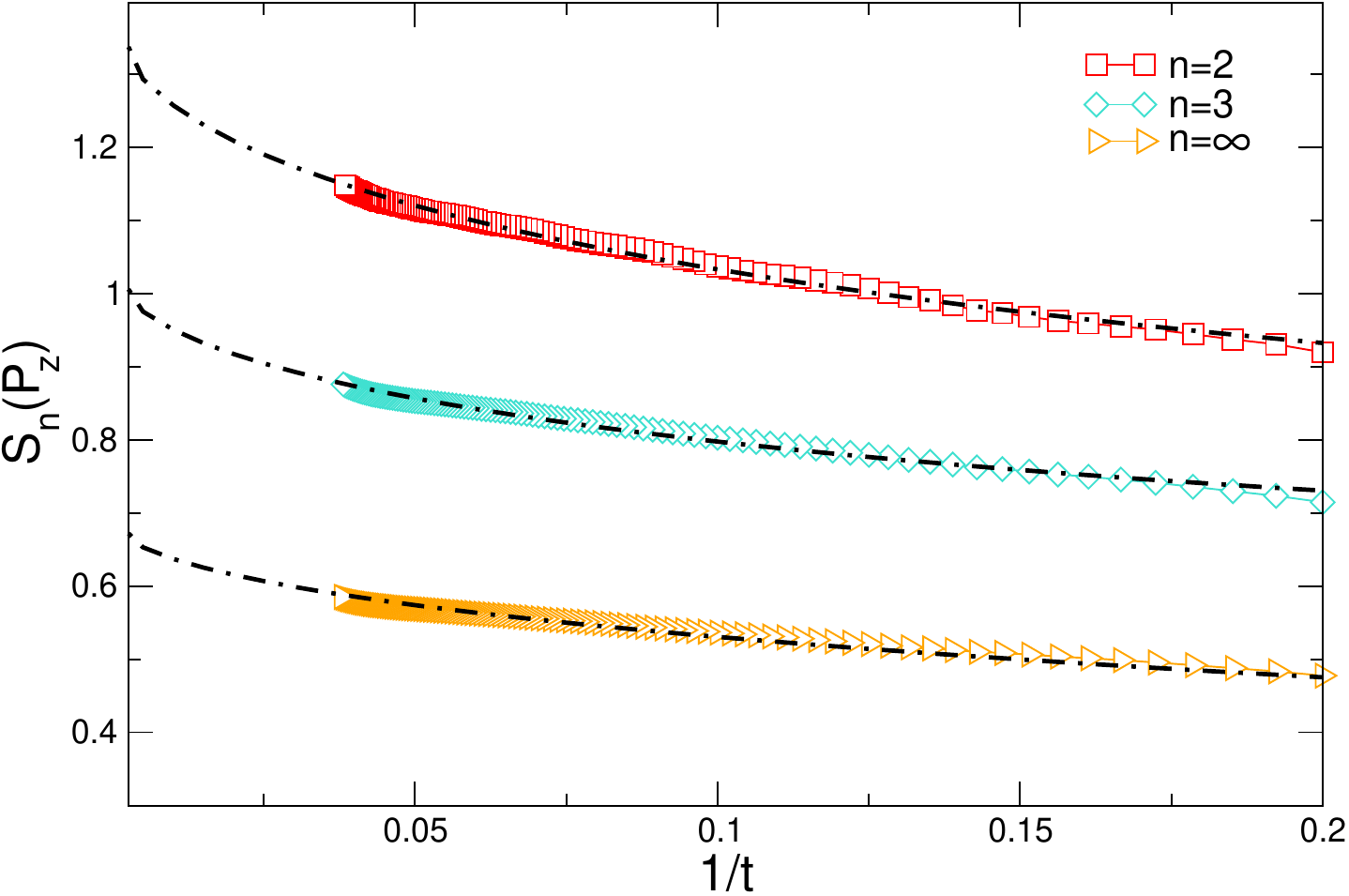}
\caption{Dynamics of the Renyi operator entanglement entropies $S_n$ for the projector operator $P_z$ in the 
	$XXZ$ spin chain with $\Delta=10$. Left panel: $tDMRG$ data for $S_n$ plotted versus time $t$. 
	The dashed-dotted line  is a fit to  $S_1=a\ln(t)+b$, with $a,b$ fitting parameters. For $n=1$ 
	the fit gives $a\approx 1/2$. The dashed lines are~\eqref{eq:renyi-simp}. 
	Right panel: Same as in the left panel using the logarithmic scale on the $x$-axis. 
	The dashed-dotted lines are fits to $S_n=n/(n-1)\ln(2)+a_1/t^{1/2}+a_2/t$, with $a_1,a_2$ fitting 
	parameters. 
\label{fig3:xxz-Pz}
}
\end{center}
\end{figure}
%

\subsection{Traceless versus  operators with nonzero trace} 
\label{sec:trace-xxz}

Let us consider the R\'enyi entropies $S_n$ of the projector operator $P_z$ in the $XXZ$ chain. 
We report $tDMRG$ data in Fig.~\ref{fig3:xxz-Pz} plotting $S_n(P_z)$ versus time for $n=1,2,3,\infty$. 
The data are for the $XXZ$ chain with $\Delta=10$. Clearly, $S_1(P_z)$ increases logarithmically with time 
(see left panel in Fig.~\ref{fig3:xxz-Pz}). The dashed-dotted line in the figure is a fit to the 
behavior $S_1=a\ln(t)+b$, with $a,b$ fitting parameters. The fit gives $a\approx 1/2$ in agreement with 
the results of Ref.~\cite{alba2021diffusion}. For $n>1$, the dynamics of $S_n$ is compatible with a saturation 
at $t\to\infty$, similar to the rule $54$ chain (see Section~\ref{sec:renyi-54}). In the right panel of 
Fig.~\ref{fig3:xxz-Pz} we plot $S_n-n/(n-1)\ln(2)$ versus $1/t$ for $n=2,3,\infty$. The dashed dotted lines are 
fits to 
\begin{equation}
	\label{eq:fit-n}
	S_n=\frac{n}{n-1}\ln(2)+ \frac{a_1}{t^{1/2}}+\frac{a_2}{t}, 
\end{equation}
with $a_1,a_2$ fitting parameters. The behavior of the corrections as $1/t^{1/2}$ is motivated by the 
results for the rule $54$ chain (see Fig.~\ref{fig1:bobenko-tr}). The dashed-dotted lines are 
obtained by fitting the data in the interval $1/t\in(0,.1]$. 
The agreement with~\eqref{eq:fit-n} is satisfactory, confirming that the data are compatible with 
$S_n(t)\to n(n-1)\ln(2)$ for $t\to\infty$. 

\begin{figure}[t]
\begin{center}
\includegraphics[width=.45\textwidth]{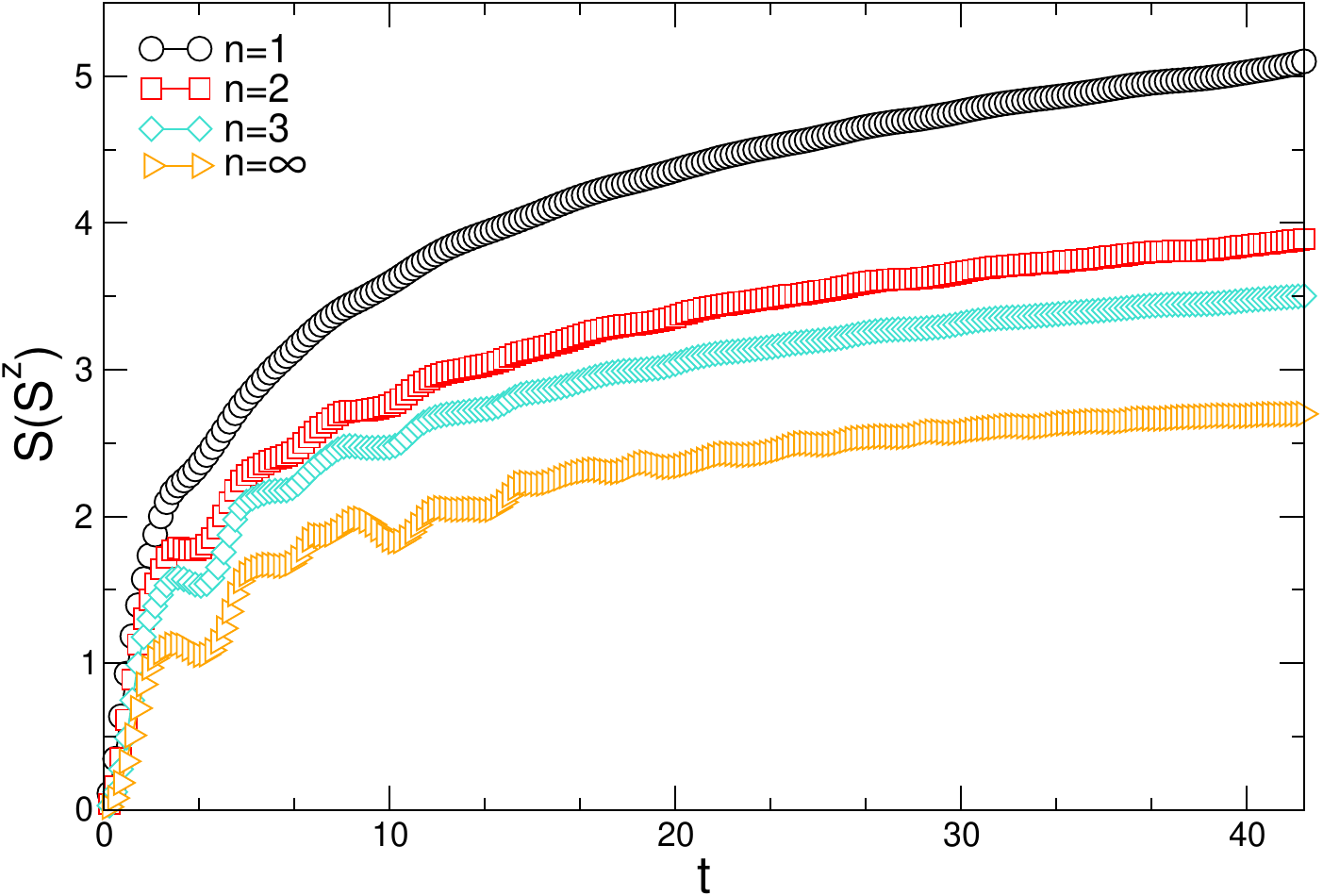}
\includegraphics[width=.45\textwidth]{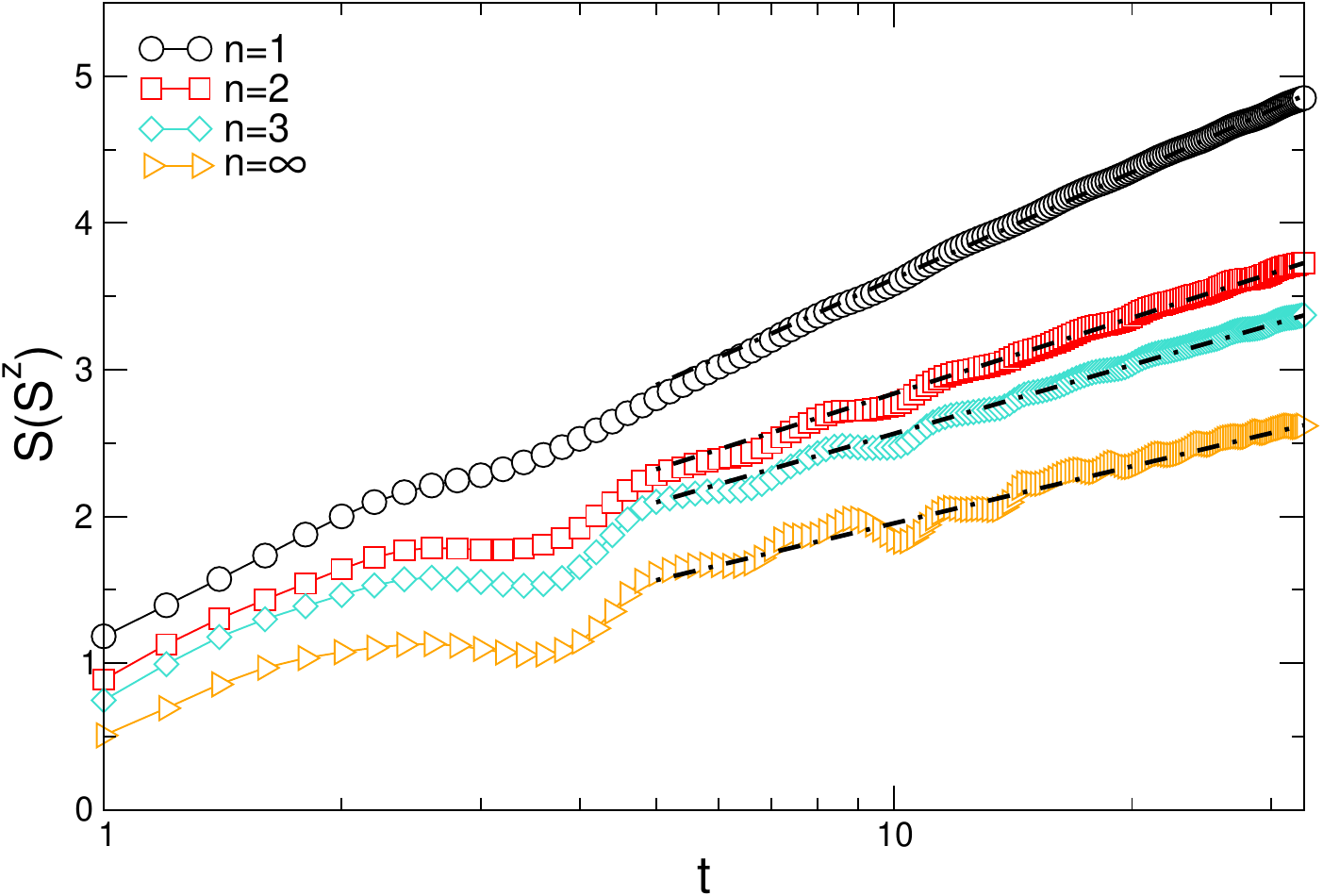}
\caption{Dynamics of the Renyi operator entanglement entropies $S_n$ for the operator $S_z$ in the 
	$XXZ$ spin chain with $\Delta=10$. Left panel: $tDMRG$ data for $S_n$ plotted versus time $t$. 
	Right panel: Same as in the left panel using the logarithmic scale on the $x$-axis.  The dash-dotted 
	lines are fits to~\eqref{eq:fit-3}.   
}
\label{fig3:xxz-Sz}
\end{center}
\end{figure}

Let us now consider the operator $S^z$. In Fig.~\ref{fig3:xxz-Sz} 
we show $tDMRG$ data for $S_n$ for the $XXZ$ chain with $\Delta=10$. In contrast with the case of $P_z$ 
the left panel of Fig.~\ref{fig3:xxz-Sz} suggests that $S_n$ grows with time for any $n$, as for the rule $54$ chain. 
This is supported in the right panel of Fig.~\ref{fig3:xxz-Sz} plotting $S_n$ versus $t$ and using a logarithmic scale on the 
$x$ axis. The dashed-dotted lines in Fig.~\ref{fig3:xxz-Sz} are fits to 
\begin{equation}
	\label{eq:fit-3}
	S_n=a\ln(t)+b+\frac{c}{t^{1/2}},
\end{equation}
with $a,b,c$ fitting parameters. By fitting the data with $t>10$ we obtain 
$a=1.05(5),0.70(5),0.66(5),0.50(5)$ for $n=1,2,3,\infty$. Interestingly, the fitted values 
of $a$ are consistent with the results obtained for the rule $54$ chain (see Section~\ref{sec:renyi-54}), 
although the time scales that are accessible are not enough to clarify whether they are 
the same. 

%
%
It is interesting to discuss the validity of the bound in Eq.~\eqref{eq:ineq}, and whether  it is saturated.  
Given an operator $\hat{\mathcal O}$ with $ITAC$ as  $\langle\hat {\mathcal O}(t)\hat{\mathcal O}(0)\rangle_{T=\infty}
\sim t^{-\alpha}$, with 
$\alpha>0$, if we assume that the bound~\eqref{eq:ineq} is saturated we obtain 
\begin{equation}
	\label{eq:pred}
	S_n\sim\frac{2\alpha n}{n-1}\ln(t). 
\end{equation}
For $\alpha=1$ and $\alpha=1/2$, which correspond to ballistic and diffusive operator spreading, Eq.~\eqref{eq:pred} 
gives $S_n\sim2n/(n-1)\ln(t)$ and $S_n\sim n/(n-1)\ln(t)$, respectively. Now, for $n=2$, Eq.~\eqref{eq:pred} implies  
that $S_2\sim 4\ln(t)$ and $S_2\sim 2\ln(t)$ for $\alpha=1$ and $\alpha=1/2$, respectively. On the other hand, from 
Fig.~\ref{fig3:xxz-Sz} one obtains the milder increase as $S_2\sim 0.8\ln(t)$. This confirms that the bound~\eqref{eq:ineq} 
is not saturated, at least for integrable systems.

\subsection{Breaking integrability}
\label{sec:int-break}

For generic nonintegrable systems, according to the membrane picture for entanglement spreading,  
$S_1$ grows linearly with time~\cite{jonay2018coarse}.  The key assumption of the membrane picture is that the 
entanglement profile $S_1(x,t)$, with $x$ denoting the position of the cut dividing subystem $A$ and $B$ (see Fig.~\ref{fig0:cartoon}), 
satisfies an  equation of the form 
$\partial_{t}S_1(x,t)=s_{\mathrm{eq}}\Gamma(\partial_x S_{1}(x,t))$, with $\Gamma$ the entropy production rate, 
and $s_{\mathrm{eq}}$ the density of entanglement entropy at equilibrium~\cite{nahum2017quantum}. 
Under quite general assumptions on $\Gamma$, 
the membrane picture predicts linear growth of the entropy. 
On the other hand, as it was stressed in Section~\ref{sec:muth}, even in nonintegrable systems it is possible to 
derive a logarithmic bound for the growth of the 
R\'enyi $OSE$ entropies for operators that are linked with a 
conserved quantity~\cite{muth2011dynamical} and exhibit a 
power-law decaying $ITAC$. 
%
\begin{figure}[t]
\begin{center}
\includegraphics[width=.45\textwidth]{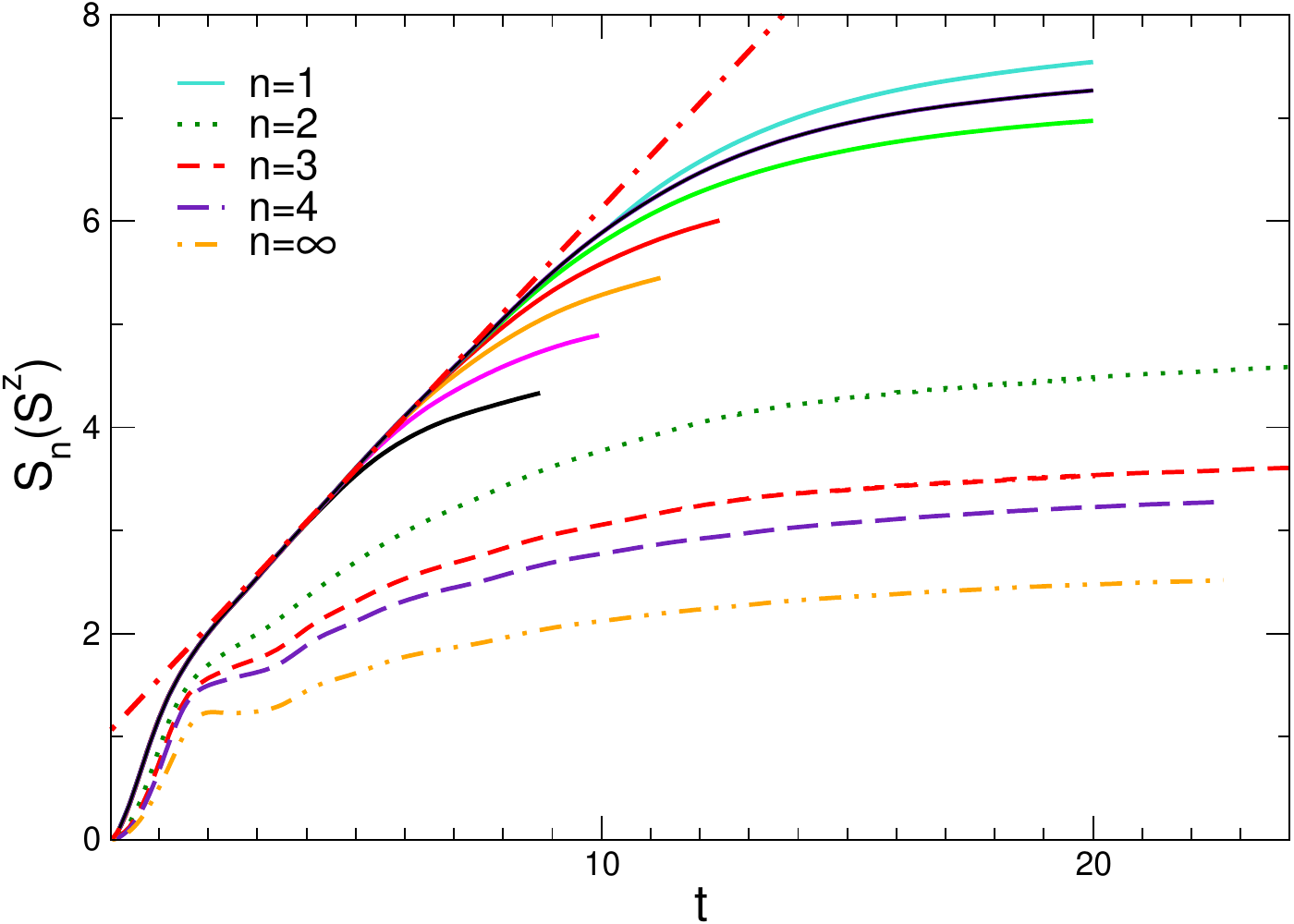}
\includegraphics[width=.45\textwidth]{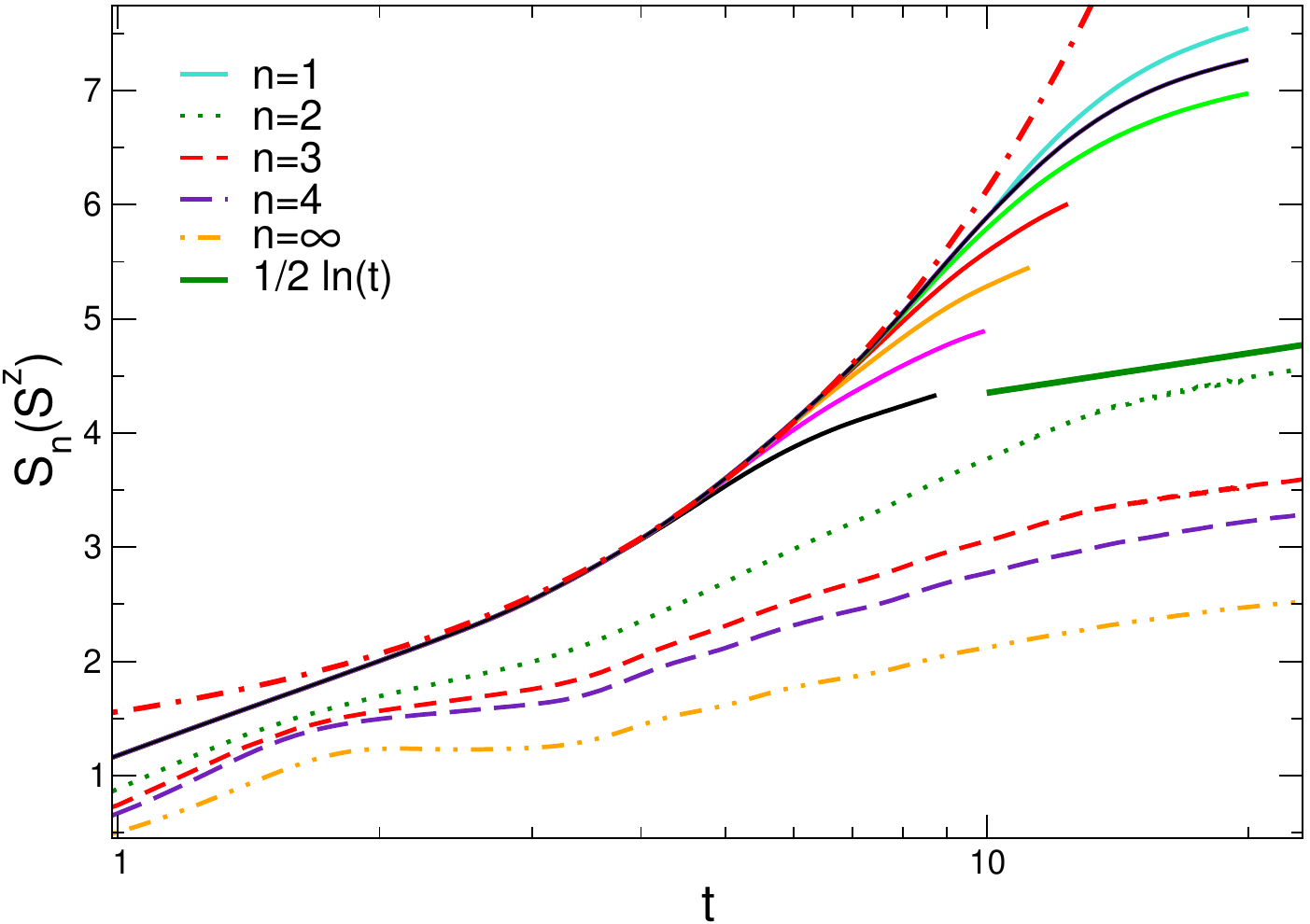}
\caption{Dynamics of the R\'enyi operator entanglement entropies $S_n$ for $S^z$ in the nonintegrable $XXZ$ chain 
	with $\Delta=3$ and $\Delta'=0.25\Delta$ (cf.~\eqref{eq:xxz-ham}). Left panel: $tDMRG$ data for $S_n$ as a function of time $t$. 
	For $n=1$ (continuous lines) the different curves correspond to different bond dimensions 
	$\chi=50,100,200,400,800,1600,2500$ (from bottom to top). The dash-dotted line is to highlight 
	linear behavior at short times. The remaining curves are $S_n$ with $n=2,3,4,\infty$. For 
	$n>1$ we show data for $\chi=1600$. Right panel: Same data as in the left panel using the logarithmic 
	scale on the $x$-axis. 
}
\label{fig5:xxz}
\end{center}
\end{figure}
%
To investigate operators spreading in the presence of integrability-breaking perturbations, in Fig.~\ref{fig5:xxz} 
we plot $S_n(S^z)$ for the deformed $XXZ$ chain (cf.~\eqref{eq:xxz-ham}) with $\Delta=3$ and $\Delta'=1/4\Delta$. For 
nonzero $\Delta'$ the Hamiltonian in Eq.~\eqref{eq:xxz-ham} is nonintegrable. In particular, by using standard 
diagnostic tools for chaotic systems, such as the distribution of energy levels, 
one can check that the spectrum of~\eqref{eq:xxz-ham} is compatible with that of a chaotic 
Hamiltonian. In Fig.~\ref{fig5:xxz} we show the von Neumann entropy for several values of the bond dimension 
$\chi=50,100,200,400,800,1600,2500$ (continuous lines from bottom to top). As already discussed in 
Ref.~\cite{alba2021diffusion}, it is challenging to obtain the 
asymptotic behavior of $S_1$ at $t\to\infty$. Precisely, the data at $t\lesssim 10$ suggest a linear 
increase (see dashed-dotted line in the Figure). However, it is not clear whether the 
bending of the curves at long times signals a change in the 
behavior of $S_1$, or it is due to truncation of the bond dimension. In Fig.~\ref{fig5:xxz} we also 
show results for $S_n$ with $n=2,3,4,\infty$. Now, the growth of $S_n$ is much weaker. 
We observe that the convergence with increasing $\chi$ is faster. This is expected because the R\'enyi entropies are 
more sensitive to the lower part of the $ES$, which is better captured even at small $\chi$. Indeed, 
in Fig.~\ref{fig5:xxz} the data for $S_n$ with $n>2$ are obtained with 
$\chi=1600$. Now, in contrast with $S_1$, $S_n$ are 
compatible with a logarithmic increase for $n>1$. Indeed, since the global magnetization is preserved 
by~\eqref{eq:xxz-ham}, even for nonzero $\Delta'$, it is reasonable to expect a power-law decaying 
$ITACC$. In turn,  this implies (cf.~\eqref{eq:ineq}) at most logarithmic growth of the $OSE$ entropies of $S^z$. 
This is also confirmed in the right panel of Fig.~\ref{fig5:xxz} where we employ a logarithmic scale on 
the $x$ axis. Notice that the prefactor of the putative logarithmic growth is $\sim 1/2\ln(t)$, i.e., 
even smaller that in the integrable case (see Fig.~\ref{fig3:xxz-Sz}). Indeed, based on the 
behavior in Fig.~\ref{fig5:xxz}, one cannot  exclude that $S_n$ saturate at long times.

%
\begin{figure}[t]
\begin{center}
\includegraphics[width=.45\textwidth]{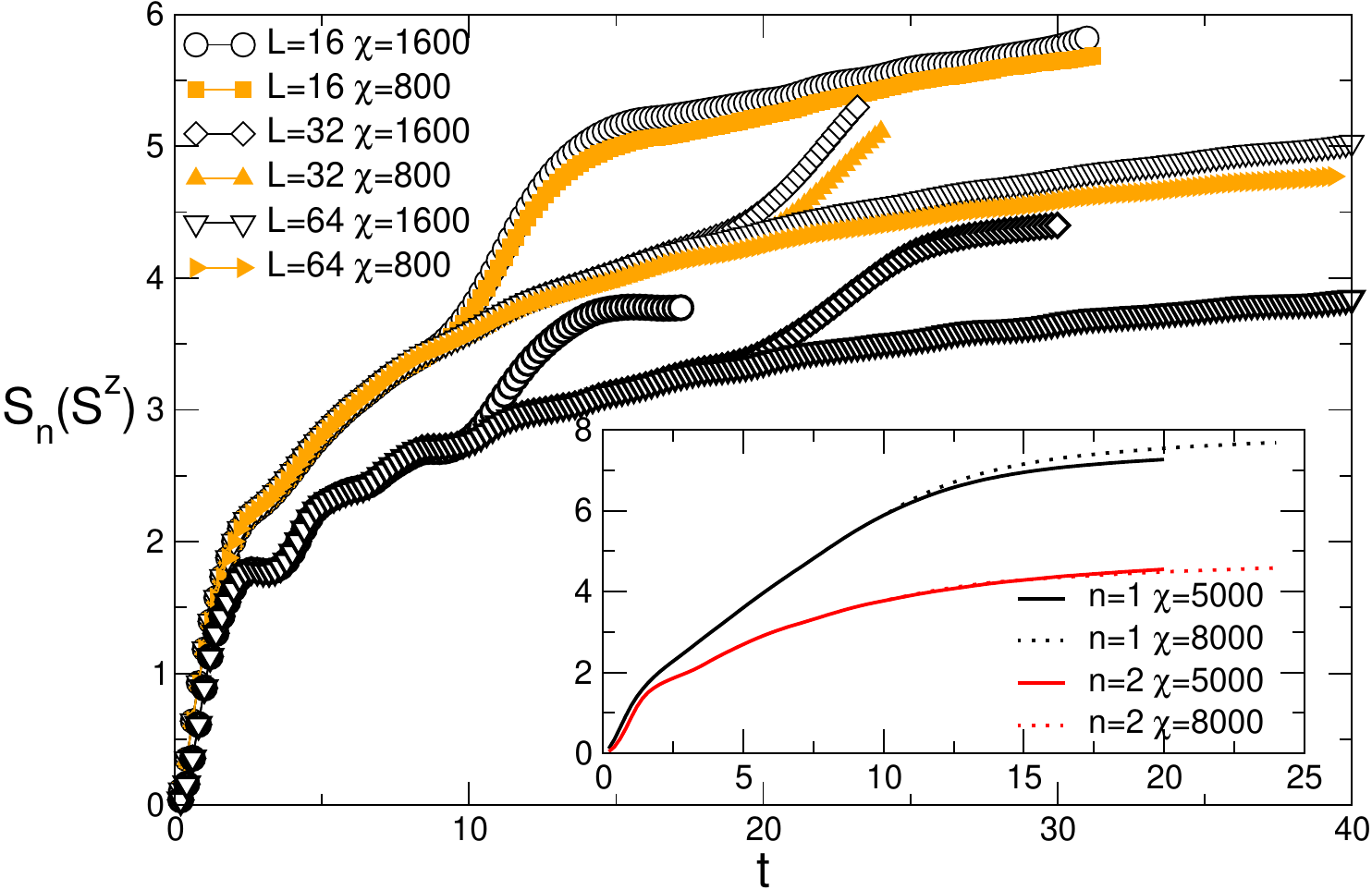}
\includegraphics[width=.45\textwidth]{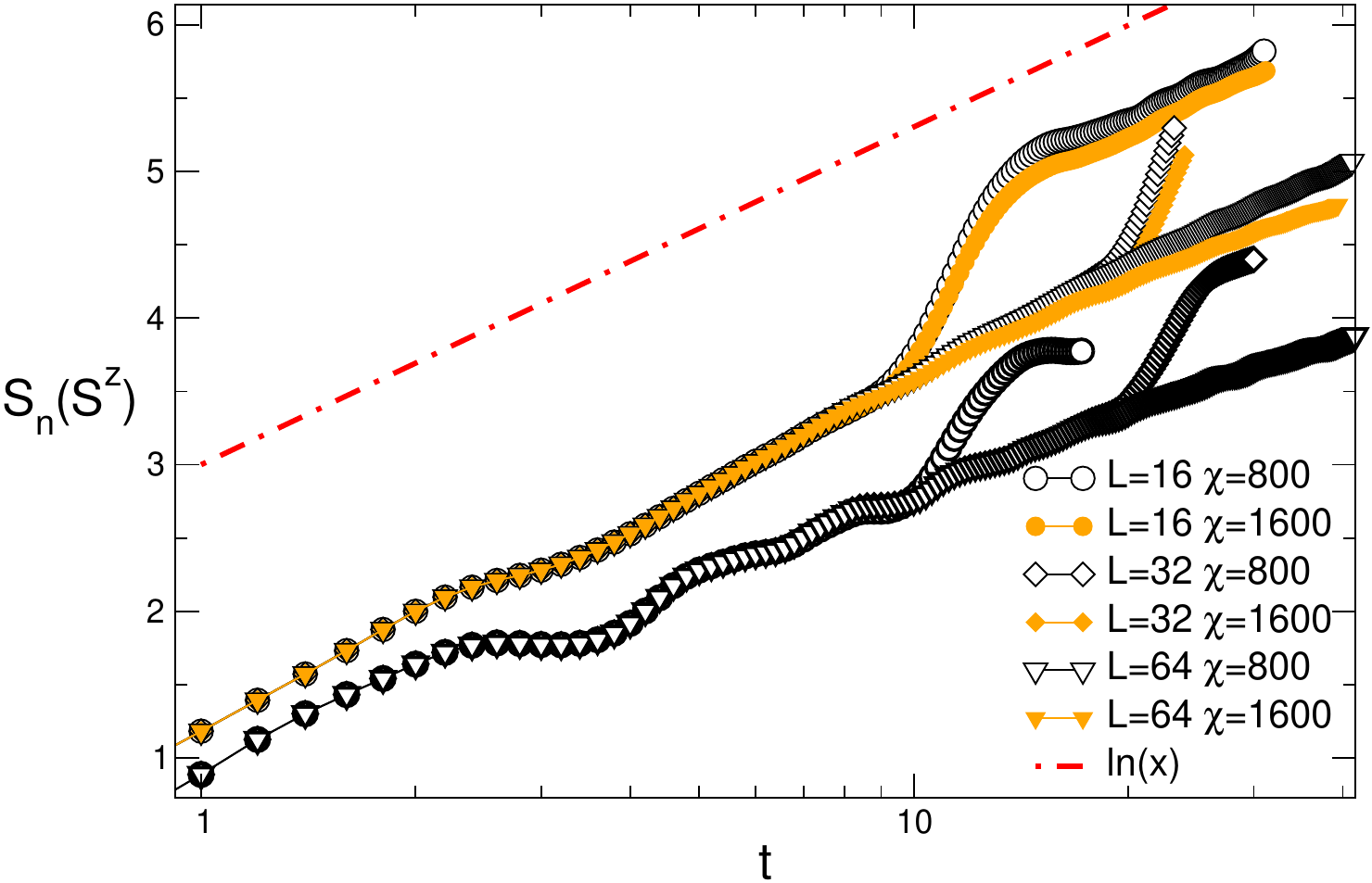}
\caption{Revival effects in the dynamics of the operator entanglement $S_n$ for $S^z$. Both panels 
	show $tDMRG$ data for the $XXZ$ chain with $\Delta=3$. Left panel: dynamics of $S_n$ for 
	a chain with $L=16,32,64$ (different symbols) and two values of the bond dimension $\chi=800,1600$. 
	We show results for $n=1$ and $n=2$. Revivals at $t\approx L/2$ are visible. The inset shows the 
	effect of integrability breaking. The curves are $tDMRG$ data for the $XXZ$ chain with $\Delta=3$ 
	and $\Delta'=0.25\Delta$. The data are for $L=16$ and $n=1,2$ with  $\chi=5000,8000$. The data show no revivals. 
	Right panel: Same data as in the main left panel using the logarithmic scale on the $x$-axis. The dash-dotted 
	line is $\ln(t)$. 
}
\label{fig6:xxz-revival}
\end{center}
\end{figure}
%

\subsection{Entanglement ``revivals'' \& integrability breaking}
\label{sec:revivals}

It is interesting to investigate the effect of the finite-size of the chain in the 
dynamics of the operator entropies. It has been shown in Ref.~\cite{modak2020entanglement} that 
in finite-size integrable chains the dynamics of the state entanglement 
 after a global quantum quench exhibits ``revivals''. The origin of revivals is that 
in integrable systems information spreads via 
the ballistic transport of quasiparticles. Moreover, quantum correlations between pairs of 
quasiparticles are preserved during their dynamics~\cite{calabrese-2005,fagotti2008evolution,alba2017entanglement}. 
The entanglement between two spatial regions is due to entangled pairs that 
are shared between the regions. Now, if the regions are embedded in a \emph{finite} chain 
the motion of the quasiparticles is periodic, implying periodicity in the contribution of the pairs 
to the entanglement between the regions. 
This explains revivals in the entanglement of a state.  Notice that  since not all the 
entangled pairs  travel at the 
same velocity, the entanglement revivals weaken with the revival time\cite{modak2020entanglement}. 
This is a weak form of quantum information scrambling~\cite{alba2019quantum} for integrable systems. 
Oppositely, it was shown in Ref.~\cite{modak2020entanglement} 
that for dynamics under nonintegrable Hamiltonians, revivals are more strongly suppressed. 
The reason is that entanglement dynamics in nonintegrable systems does not happen via quasiparticle spreading. 
Instead, information is quickly dispersed in the global degrees 
of freedom by the dynamics. 

Here we show that revivals occur in the operator entanglement spreading in finite chains, 
although they have a different origin. In the following we focus on the 
dynamics of traceless operators in the deformed $XXZ$ chain~\eqref{eq:xxz-ham}.  
Our $tDMRG$ are reported in Fig.~\ref{fig6:xxz-revival}. In the left panel of Fig.~\ref{fig6:xxz-revival} 
we plot $S_n(S^z)$  for the $XXZ$ chain with $\Delta=3$ ($\Delta'=0$ in~\eqref{eq:xxz-ham}). 
We focus on $n=1$ and $n=2$. We plot 
with different symbols (circles, diamonds, and triangles) the dynamics in chains of different 
sizes $L=16,32,64$. For $n=1$ the empty symbols correspond to $\chi=1600$, whereas the full 
symbols are the results for $\chi=800$. For $n=2$ we only show the results for $\chi=1600$, because 
it is sufficient to ensure convergence. 
We plot the same data in the right panel in Fig.~\ref{fig6:xxz-revival} using a logarithmic scale on the 
$x$ axis. The revivals of the entropies are clearly visible. Interestingly, the prefactor of the logarithmic 
growth after the revival remains the same. 
Notice that the velocity of the quasiparticles, and hence the 
time of the revivals, could be determined via the Thermodynamic Bethe Ansatz 
approach~\cite{gopalakrishnan2018hydrodynamics}. 
Let us now discuss the mechanism underlying these revivals. First, as it is clear from the case of the rule $54$ chain,  
(see Section~\ref{sec:rule-54}), the growth of the entropies is not due to the propagation of entangled pairs. 
Instead, the growth of the $OSE$ is due to the fact that the information needed to detect the position of the 
solitons generated by the operator insertion at the center grows with time. 
Now, the ``kicks'' at the revivals can be interpreted as the extra information needed to decide whether the solitons emitted at the center 
were reflected at the boundaries or not. 
Let us now discuss the effect of breaking integrability. We show $tDMRG$ data for the nonintegrable $XXZ$ chain with $L=16$ 
with $\Delta=3$ and $\Delta'=\Delta/4$ in the inset of Fig.~\ref{fig6:xxz-revival}. The data do not show any sign of 
entanglement revivals. This is consistent with the fact that in nonintegrable systems 
information is not spread by propagation of quasiparticles.

\section{Conclusions}
\label{sec:concl} 

We numerically investigated the dynamics of the R\'enyi Operator Space Entanglement ($OSE$) entropies in one dimensional 
systems. We focused on both integrable and nonintegrable systems. 
We first discussed the paradigmatic case of the rule $54$ chain. 
We showed that the R\'enyi operator entanglement entropies $S_n$ of traceless operators grow logarithmically 
with time for any $n$. The prefactor of the logarithmic growth depends on $n$ in a nontrivial manner. 
On the other hand, for operators that have nonzero trace, $S_n$ grow logarithmically 
only for $n=1$. For $n>1$, $S_n$ saturate in the limit $t\to\infty$, and the saturation values are 
determined by the operator trace. 
Moreover, our results suggest that the full entanglement spectrum  of operators with 
nonzero trace can be reconstructed from the entanglement spectrum of their traceless part. 
Interestingly, the scenario is similar for the $XXZ$ chain, which is a prototypical Bethe ansatz 
integrable model. Specifically, the entropies of operators with nonzero trace grow logarithmically for $n=1$, 
whereas they saturate for $n>1$. For traceless operators, $S_n$ 
grows logarithmically for any $n$. The prefactors of the logarithmic growth are compatible with the ones 
numerically obtained for the rule $54$ chain. 

We next discussed operator spreading in nonintegrable dynamics. As it was shown in Ref.~\cite{alba2021diffusion}, 
it is challenging to extract the asymptotic behavior of $S_1$ at long times. On the other hand, 
the growth of $S_n$ for $n>1$ is milder as compared with $n=1$, and it is compatible with a logarithmic growth, 
at least for the models that we considered. 
Finally, we showed that in finite integrable systems, $S_n$ 
exhibit revivals. After the revivals $S_n$ continue 
growing logarithmically with time. Interestingly, upon switching on  integrability-breaking 
interactions, revivals disappear, signaling the absence of well-defined quasiparticles. 

There are several directions for future research. First, it is important to clarify the scaling of the 
$OSE$ entropies in nonintegrable systems. This is a challenging task, as already pointed out in Ref.~\cite{alba2021diffusion}, 
because of strong finite-time corrections. Furthermore, it would be important to clarify the mechanism 
underlying the growth of $OSE$ in integrable systems. Precisely, the fact that the prefactor of the logarithmic growth of the 
entropies is the same for both the rule $54$ chain and the $XXZ$ chain suggests that 
it could be universal. It would be useful  to numerically extract the 
prefactor of the logarithmic growth by using the $MPO$ representation of local 
operators~\cite{klobas2019time,alba2019operator,foligno2024entanglement} in the rule $54$ chain. 
This will allow to clarify the relationship between $OSE$ growth and transport properties. 
Notice that a deep relationship between operator growth and transport has been  already established 
in Ref.~\cite{wellnitz2022rise} for some one-dimensional systems subject to dephasing. Moreover, it would be 
interesting to investigate the $OSE$ growth in random unitary circuits with $U(1)$ conservation laws, 
by employing the results of Refs.~\cite{rakovzsky2019sub,huang2020dynamics}. 
On the analytical side, one could try to 
explore the relationship~\cite{dowling2023scrambling} between $OSE$ and Out-of-Time-Order-Correlators ($OTOC$). 
Finally, having access to the asymptotic behavior of the R\'enyi entropies would allow to 
reconstruct the structure of the operator entanglement spectrum. 

\section*{Acknowledgements}

\emph{This paper is dedicated to the memory of Marko Medenjak who left us before his time.  
 In his brief journey Marko has left many important contributions that helped shape my interest in 
 operator spreading. More importantly, our discussions stimulated many interesting questions, 
 some of which are explored in this work. }

This study was carried out within the National Centre on HPC, Big Data and Quantum Computing - SPOKE 10 (Quantum Computing) and received funding from the European Union Next-GenerationEU - National Recovery and Resilience Plan (NRRP) – MISSION 4 COMPONENT 2, INVESTMENT N. 1.4 – CUP N. I53C22000690001. 
This work has been supported by the project ``Artificially devised many-body quantum dynamics in low dimensions - ManyQLowD'' funded by the MIUR Progetti di Ricerca 
di Rilevante Interesse Nazionale (PRIN) Bando 2022 - grant 2022R35ZBF.  

\appendix

\section*{References}
\bibliographystyle{iopart-num.bst}
\bibliography{bibliography}

\end{document}